\documentclass{article}
\usepackage[utf8]{inputenc}
\usepackage{amsmath,amssymb}
\usepackage{graphicx}
\usepackage{booktabs}
\usepackage{caption}
\usepackage{subcaption}
\usepackage{float}
\usepackage{fancyhdr}
\usepackage{geometry}
\usepackage{setspace}
\usepackage{hyperref}
\hypersetup{
    colorlinks=true,
    linkcolor=blue,
    citecolor=blue,
    urlcolor=blue,
}
\usepackage{cleveref}

\title{When AI Deceives: A Natural Experiment on the Causal Effects of Perceived Deception on Player Ratings in RPGs}
\author{Shudong YANG$^1$\\
\small $^1$School of Art and Design, Dalian Jiaotong University, China}
\date{}

\begin{document}

\maketitle

\begin{abstract}
\textbf{Objective:} AI-driven deception mechanisms are increasingly prevalent in digital games, yet the direction and magnitude of their effects on player experience remain contested. Existing research has not sufficiently disentangled designer-intended deception intensity from players' actual perception of deception, and most prior work relies on low-ecological-validity experiments or cross-sectional surveys. The present study aims to independently examine the causal effects of design deception intensity (DDI) and player deception awareness (PDA) on player ratings within a naturalistic gaming environment, and to investigate the moderating role of player experience. \\
\textbf{Methods:} Leveraging the 54 version updates of \emph{Baldur's Gate 3} between 2019 and 2025 as a quasi-natural experiment, it collected all English-language Steam reviews posted within 1 to 28 days following each update (N = 160,835), and constructed a player--version two-way fixed effects panel dataset. DDI was coded by human annotators based on patch notes (ICC = 0.77); PDA was extracted and aggregated from review texts using a fine-tuned BERT classifier (F1 = 0.84). The model incorporated both player and version fixed effects, complemented by five robustness checks including subsample partitioning, lagged variables, and placebo tests. \\
\textbf{Results:} PDA exerts a monotonic negative effect on positive review rates: within the observed PDA range (2\%--6\%), the net loss in review valence is approximately 0.4 percentage points ($\beta = -0.0969$, $p < 0.001$), with a negative quadratic term that falsifies the inverted-U hypothesis of moderate perception optimality. DDI exhibits a U-shaped effect with an inflection point at a relatively low intensity (DDI = 1.67), although the upward trend on the right branch is primarily driven by contemporaneous new content bundled with high-intensity updates. Player experience produces a strong desensitization effect: novice players exhibit a pronounced decline in ratings when perceiving deception ($\beta = -0.107$), whereas veteran players remain largely unaffected ($\beta = 0.031$, $p = 0.346$). All five robustness checks consistently support these conclusions. \\
\textbf{Conclusion:} Any degree of deception awareness undermines player evaluations, while the positive manifestation of design intensity depends on content-confounding effects. It recommend that game designers prioritize reducing deception exposure for novice players, or bundle deceptive mechanisms with high-value content releases. This study validates the methodological feasibility of version-update-based natural experiments in games user research, and provides empirical grounding for responsible AI game design.
\end{abstract}

\section{Introduction}

AI-driven deception mechanisms are becoming increasingly common in digital games~\cite{yannakakis2025}, ranging from narrative-level misleading dialogue and hidden plotlines to mechanical-level deceptive behaviours by non-player characters (NPCs) and disguised randomness~\cite{starace2026,mareddy2025}. Such designs are intended to enhance challenge, immersion, and replayability; however, their potential negative consequences, players feeling manipulated, experiencing eroded trust, and reporting diminished enjoyment~\cite{poivet2025} have fuelled ongoing debate within game research.

On the one hand, when players explicitly recognise that they have been deceived by AI, their satisfaction, perceived fairness, and willingness to continue playing may decline. This aligns with the predictions of Expectancy Violation Theory: once players' psychological expectation that game AI should be fair and transparent is breached, negative emotional responses are likely to ensue~\cite{poivet2023}. On the other hand, player comments on Steam suggest that certain forms of deception, such as plot twists and hidden clues, may be received neutrally or even positively~\cite{yin2024}, indicating that the impact of deception is not univocally negative but may exhibit non‑linear or moderated patterns.

Three notable research gaps merit attention in the existing literature. First, there is a persistent confounding between designer‑intended deception intensity and players' actual perception of deception. Some studies manipulate only the presence or absence of deception, or directly ask players whether they felt deceived, without adequately separating objective design intensity from subjective perception. This leaves open the questions of whether players genuinely perceive the complex deception mechanisms carefully crafted by designers, and how design intensity and player perception independently influence experience. Second, the moderating role of player experience has not been sufficiently examined~\cite{kocielnik2025}. Novice and veteran players differ markedly in cognitive schemas, expectation levels, and tolerance for deception, yet how experience moderates the effects of deception remains under‑explored. Third, methodological limitations persist: laboratory settings fail to replicate the complex social interactions and sustained immersion of real‑world gameplay, raising concerns about ecological validity~\cite{allen2024}; meanwhile, questionnaire‑based surveys rely on post‑hoc self‑reports, which are susceptible to recall bias and social desirability~\cite{jeong2018}. Both paradigms encounter difficulties in achieving low‑cost, large‑sample causal inference.

To address these limitations, the present study proposes and validates a natural‑experimental design grounded in game version updates. This design is intended to circumvent the artificiality of laboratory settings while overcoming the endogeneity problems inherent in cross‑sectional surveys. Large‑scale role‑playing games (RPGs) such as \emph{Baldur's Gate 3} undergo continuous version updates, the majority of which incorporate deceptive design elements to varying degrees. The timing of these updates is determined by developers and is exogenous to players, thus constituting a naturally occurring intervention. By collecting player reviews within a defined post‑update window and exploiting panel data structures to control for individual heterogeneity and time trends, this approach enables causal inference within an authentic gaming context at relatively low cost.

Accordingly, this study leverages the 54 version updates of \emph{Baldur's Gate 3} between 2019 and 2026 as natural experimental events. It collect all English‑language Steam reviews posted within 1 to 28 days following each update (N = 160,835) and construct a player--version two‑way fixed effects panel dataset. The core objectives are fourfold: 1) to disentangle Design Deception Intensity (DDI) from Player Deception Awareness (PDA) and examine their respective main effects on player ratings; 2) to test whether PDA exhibits non‑linear effects such as an inverted‑U shape; 3) to investigate whether player experience moderates these effects; and 4) to validate the feasibility of the version‑update‑based natural experiment method for games user research. The findings will provide empirical evidence for responsible AI game design and offer a methodological reference for innovation in game research.

Following the IMRaD structure, the remainder of this paper is organised as follows: Section~2 reviews related work; Section~3 describes the data and methods; Section~4 presents the core results; Section~5 discusses the findings, contributions, and limitations; and Section~6 concludes the study.

\section{Related Work}

\subsection{Theoretical Basis of Game AI Deception}

Game AI deception refers to algorithmically driven behaviours of non‑player characters (NPCs) or systems, excluding game bugs, that aim to influence players' decisions, cognition, or emotional states by concealing truthful information, providing false cues, or disrupting conventional patterns~\cite{starace2026}. Unlike everyday interpersonal deception, deception within digital game contexts is often designed as a legitimate foul, simultaneously violating players' default expectations while being justified by developers as a necessary means to enhance challenge and immersion~\cite{gualeni2021}.

Two theoretical perspectives are most frequently invoked to account for player responses to AI deception. The first is Expectancy Violation Theory (EVT). Within the gaming context, this theory posits that players hold prior expectations regarding NPC or system behaviours, and that deviations from these expectations trigger cognitive arousal and evaluative revision~\cite{burgoon2016}. In the specific context of game AI deception, players typically anticipate that the game world should adhere to predictable and fair rules. Once AI deception occurs, it constitutes a negative expectancy violation, potentially eliciting negative emotions such as anger, frustration, or diminished trust~\cite{poivet2025}. However, EVT also predicts that the consequences of expectancy violation depend on the reward valence of the violator; if the violation is interpreted as benevolent or amusing, positive outcomes may emerge~\cite{yin2024}. This provides a theoretical foundation for the view that moderate deception might be acceptable, although empirical evidence remains inconsistent.

The second relevant theoretical lens is Psychological Contract Theory. In gaming contexts, an informal psychological contract forms between players and the game, encompassing implicit provisions such as fairness, transparency, and the comprehensibility of AI. When AI deception becomes frequent or severe, this psychological contract is breached, leading players to reduce emotional investment, lower their evaluations, or even abandon the game~\cite{johnson2012}. This theory underscores the importance of player perception: even if designers consider the deception to be minor, the contract is violated as long as players perceive themselves as having been deceived.

Overall, existing theories that explain player responses to AI deception predict that the effects are jointly determined by design intensity, player perception, and individual characteristics such as experience and trust propensity~\cite{leiser2025}. Nevertheless, the theoretical distinction between perception and design has yet to be systematically operationalised in empirical work, leaving room for theoretical refinement that the present study aims to address.

\subsection{The Empirical Research in Game Science}

Existing empirical research in game science can be categorised into three predominant methodological approaches.

The first is \textbf{laboratory experiments}. In such studies, researchers typically have participants interact with a game in a controlled environment and subsequently measure indicators such as satisfaction, perceived fairness, and immersion. For instance, Denisova and Cairns (2019) conducted a laboratory experiment in which participants played a game that did not actually feature difficulty adaptation; however, some participants were informed that such adaptation was present. Using immersion questionnaires, they found that the mere provision of this information significantly enhanced players' immersion~\cite{denisova2019}. While laboratory experiments offer high internal validity, they are often constrained by small sample sizes, typically fewer than 200 participants, and the artificiality of the tasks, which limits players' long-term immersion and emotional investment. Consequently, their ecological validity is frequently questioned~\cite{subramanian2023}.

The second approach is \textbf{questionnaire surveys}. Researchers administer self-report questionnaires to players, asking about their perceptions of AI deception and their satisfaction levels. For example, Luis Duarte (2017) surveyed 196 players to assess the prevalence and characteristics of deception mechanisms in contemporary digital games and their relationship to game mechanics. Although questionnaire surveys can reach larger samples, they are susceptible to recall bias and social desirability bias, and they struggle to establish causal relationships~\cite{podsakoff2003}.

The third approach is review text analysis. Some studies employ natural language processing techniques to mine deception-related topics from Steam or forum reviews. For example, Lin and colleagues (2019) conducted natural language processing and classification analyses on nearly 11 million review texts from 6,224 games on the Steam platform, revealing differences between game reviews and mobile app reviews in terms of content, length, and informative value~\cite{lin2018}. The strengths of this approach lie in its use of authentic, large-scale data; however, it is limited by dictionary coverage and semantic understanding capabilities, and most analyses remain cross‑sectional and correlational, making it difficult to control for individual heterogeneity.

Taken together, the existing empirical literature exhibits four limitations: 1) objective design intensity and subjective player perception have not been separated; most studies equate the presence of deception with players' perception of it, overlooking their potential divergence; 2) there is a lack of systematic examination of the moderating role of player experience, with the differences between novice and veteran players rarely quantified; 3) causal inference is weak, laboratory experiments suffer from low ecological validity, while non‑experimental approaches struggle with endogeneity; and 4) there is a scarcity of long‑term panel data that can track the same players' dynamic responses across multiple version updates.

\subsection{Application of Natural Experiment in Game Research}

To address the aforementioned limitations, some studies have begun to adopt quasi‑experimental and panel data approaches. Natural experiments leverage exogenous institutional or technological changes as treatment variables, enabling causal identification in non‑experimental settings~\cite{egami2024}. In the domain of game research, version updates, A/B tests, and phased rollouts can all serve as natural experimental events. For instance, Claypool and colleagues (2017) utilised patch updates in League of Legends as exogenous technological changes, analysing large‑scale player data before and after the patches to reveal their impact on game balance metrics such as champion win rates~\cite{claypool2017}. Similarly, Chen and colleagues (2023) exploited a system vulnerability on an online game platform that led to player misconduct, along with subsequent recovery measures, as a quasi‑natural experiment. Using panel data methods, they analysed the causal effects of the vulnerability and different recovery strategies on players' playtime and spending behaviour~\cite{chen2023}. These studies provide preliminary evidence for the technical feasibility of treating version updates as natural experiments.

Panel data methods further enhance the reliability of causal inference in game science research. The two‑way fixed effects (TWFE) model absorbs time‑invariant player‑specific heterogeneity, such as inherent preferences, through player fixed effects, and absorbs common shocks, such as seasonal trends or public opinion shifts, through time or version fixed effects, thereby yielding cleaner causal estimates~\cite{kang2024}. Moreover, in game research, clustering standard errors or accounting for autocorrelation structures at the player level can address serial correlation, further improving the accuracy of statistical inference.

Nevertheless, the application of natural experiments combined with panel data methods to the study of AI deception remains an unexplored research gap. No prior study has leveraged version update events to identify the causal effects of player deception awareness on ratings, nor has any attempted to disentangle design intensity from perceived intensity in this manner.

\subsection{Research Gap}

Although the studies reviewed above provide an important foundation for understanding AI deception in games, a comprehensive synthesis reveals four clear research gaps in the existing literature.

\textbf{First, there is a gap concerning the confounding of design intensity and player perception}. Prior studies have rarely, if ever, systematically disentangled designer‑intended deception intensity (DDI) from players' actual perceived deception (PDA). Laboratory experiments typically manipulate a simple binary presence/absence of deception, whereas questionnaire surveys directly ask whether players felt deceived, implicitly assuming the two are equivalent. However, patch notes may contain numerous deceptive design elements that players do not necessarily articulate in their reviews; conversely, even low‑intensity deception, if poorly implemented, may provoke strong complaints. The absence of independent quantification of DDI and PDA leaves unanswered the core question of how design intensity and player perception respectively influence experience.

\textbf{Second, there is an evidential gap regarding non‑linear effects and moderating mechanisms}. Although Expectancy Violation Theory suggests that deception effects may exhibit thresholds or inverted‑U patterns, empirical evidence remains inconclusive. Most existing studies rely on binary comparisons of deception presence versus absence, which precludes the testing of non‑linear relationships. Moreover, player experience, a critical moderating variable, has rarely been incorporated into analytical frameworks. Novice and veteran players differ systematically in cognitive schemas, expectation levels, and tolerance thresholds~\cite{küchelmann2026,michalco2015}, yet it remains unclear how experience alters the direction and magnitude of deception effects.

\textbf{Third, there is a methodological gap in causal identification}. Although laboratory experiments offer high internal validity, their small sample sizes and artificial contexts limit the external validity of findings~\cite{levitt2007}. Questionnaire surveys and cross‑sectional text analyses, while capable of covering large‑scale authentic data, struggle to control for individual heterogeneity and reverse causality~\cite{savitz2022}. The field lacks a methodological paradigm that enables low‑cost, high‑credibility causal inference within authentic gaming environments. The potential of game version updates as natural experimental events has yet to be exploited for deception research.

\textbf{Fourth, there is a gap in long‑term dynamic analysis}. The vast majority of studies employ cross‑sectional designs, observing a single time point or a single version. This neglects the dynamic changes in player perception and evaluation across different version updates, and fails to control for time‑invariant player characteristics such as personal preferences and linguistic expression styles. Panel data methods are rarely applied in game research, and there is a particular lack of longitudinal evidence tracking the same cohort of players' evaluative changes across multiple version updates.

These four gaps are interrelated and together point to an unresolved core question: In authentic gaming environments, how do designer‑intended deception intensity and player‑perceived deception independently and jointly influence player evaluations? And does this influence vary with player experience?

\section{Data and Methods}

\subsection{Data Sources and Natural Experimental Design}

The empirical data for this study are drawn from all English‑language player reviews of \emph{Baldur's Gate 3} (App ID: 1086940) on the Steam platform, as well as the 54 official version patch notes released for the game between June 2019 and November 2025. The selection of \emph{Baldur's Gate 3} as the case study is motivated by three considerations. First, the game's six‑year continuous update cycle incorporates a rich array of AI deception elements, including NPC betrayals, hidden plotlines, and disguised randomness, with patch notes that explicitly document deceptive designs, facilitating systematic coding. Second, Steam's review system provides precise timestamps, playtime duration, upvote/downvote labels, and other structured metadata, which are well‑suited for constructing panel data. Third, the game maintains a large and active player community, with a sufficient volume of reviews to support fine‑grained version‑level analyses.

The core design of the natural experiment exploits the exogenous timing of game version updates as "intervention" events. The release date of each update is determined by developers and is exogenous to players; players cannot predict or systematically select the timing of their review postings relative to the update. Therefore, taking the update date as the starting point (denoted as $t_0$), collecting player reviews posted within a defined post‑update window approximates quasi‑experimental data analogous to a post‑intervention observation period. In this study, the period from day 1 to day 28 following each update is defined as the effect window (post\_window), focusing on the immediate impact of updates on player evaluations. The upper bound of 28 days was chosen because preliminary data analysis indicated that the majority of reviews for most versions are posted within the first 30 days, and longer windows risk introducing interference from other in‑game events or external factors. Reviews posted before the update (pre\_window) were not used in this study, as preliminary analysis revealed that pre‑update reviews typically refer to content from older versions and cannot be cleanly attributed to the deception design of the current version.

All player review data used in this study are publicly available on the Steam platform, and all reviews are voluntarily published by players. All player IDs were hashed and anonymised by the researchers, and no personally identifiable information was collected or stored. According to the evaluation by the Institutional Review Board at the researchers' home institution, this study constitutes a secondary analysis of publicly available, non‑sensitive data and qualifies for exemption from full ethical review.

\subsection{Sample Screening and Preprocessing}

A total of 242,169 English-language reviews were downloaded from the Steam platform. To ensure analytical quality, a multi-step screening procedure was implemented. First, the Python \texttt{langdetect} library was applied to all reviews for language detection, retaining only English-language reviews to eliminate cross-linguistic semantic variations that might interfere with text analysis. Second, reviews with fewer than five words were excluded, as such extremely short comments typically lack substantive content and do not reliably indicate whether deception perception is involved. Third, based on version window matching rules, each review was associated with a specific version according to its publication date and the implicit version identifier (matched\_version) inferred from the review content; only reviews that were successfully matched to the post\_window of a given version were retained. Reviews that failed date parsing or fell outside any valid version window were excluded.

To construct a clean panel structure, a further de-duplication step was performed at the player--version level. Within the same version window, the same player might post multiple reviews. Given that these reviews exhibit temporal dependencies and potentially redundant content, only the most recent review from each player within each version window was retained. The final valid sample comprised 160,835 reviews, covering 49 analytical version windows; some versions were merged due to insufficient review volume or overlapping windows. The mean positive review rate of the sample was 95.63\%, consistent with the general tendency of highly rated popular games on the Steam platform.

\subsection{Variable Measurement}

The dependent variable is players' immediate evaluation of the game version, operationalised as the voted\_up label in each review (1 for positive, 0 for negative). This variable directly reflects player satisfaction and willingness to recommend. In robustness checks, the dependent variable is replaced with sentiment scores computed using the VADER sentiment analysis tool, ranging from --1 to 1, to test whether the conclusions are sensitive to the operationalisation of the outcome.

One of the core independent variables is Design Deception Intensity (DDI). This variable is derived from structured coding of the official patch notes. Three coders independently read the patch notes for each update and identified design elements involving AI deception, including but not limited to: NPCs deliberately providing false information, hidden quest trigger conditions, non‑randomisation of ostensibly random outcomes, and deception in plot twists. Each update was rated on a scale from 0 (no deception) to 5 (dense/high‑intensity deception), with intermediate levels defined as 1 (very minor), 2 (a small amount), 3 (moderate), and 4 (substantial). After two rounds of iterative coding, the inter‑coder agreement reached ICC = 0.77 (95\% CI: [0.71, 0.82]), and the median of the three coders' ratings was taken as the DDI value for each version. DDI ranges from 0.00 to 4.00, with the distribution concentrated around 0.00 (no deception), 2.00--3.33 (moderate intensity), and 4.00 (high intensity).

The second core independent variable is Player Deception Awareness (PDA). PDA is defined as the proportion of reviews within a given version window that are classified by a BERT classifier as explicitly mentioning or expressing awareness of AI deception, relative to the total number of reviews in that window. Thus, PDA is a continuous variable between 0 and 1, with higher values indicating a greater collective perception of deception among players for that version. The computation of PDA relies on a binary BERT classification model, the construction of which is detailed in the following subsection.

With respect to moderators and covariates, player experience is operationalised as the natural logarithm of the cumulative playtime at the time of review (playtime\_forever), denoted as log\_playtime, to mitigate the right skew of the raw distribution. Review length (word\_count) is included as a proxy for textual complexity, serving to control for the potential confound that longer reviews may be more likely to discuss deception. In addition, the model incorporates version fixed effects (matched\_version) to absorb all unobservable version‑level factors that do not vary across players within a version, such as update scale, contemporaneous marketing campaigns, and seasonal effects. It is worth noting that because the panel structure is at the player--version level rather than a strict time series, it do not include a time trend term; instead, version fixed effects are used to control for cross‑sectional differences across versions.

\subsection{Construction of Deception Perception Classifier (BERT)}

Constructing a high‑accuracy deception perception classifier is a prerequisite for computing PDA. It adopted a three‑stage process: seed dictionary construction guided by theory, large language model‑assisted annotation, and BERT fine‑tuning.

The first stage involved the construction and refinement of a seed dictionary. Based on theoretical literature on deception and high‑frequency terms appearing in \emph{Baldur's Gate 3} community discussions, it manually compiled an initial set of approximately 50 seed words, such as "lie," "deceive," "betray," "hidden," "trick," and "manipulate." These seed words were then fed into a large language model, with instructions to expand semantically similar keywords and phrases from game review texts. After manual deduplication and relevance filtering of the model's output candidates, an initial dictionary containing 361 words was formed. Subsequently, it tested matching performance on a small sample, removing low‑frequency, ambiguous, or falsely detected terms, and ultimately retained an optimised dictionary of 163 core keywords. This dictionary covers multiple semantic dimensions, including direct deception (e.g., "lie," "deceive"), betrayal (e.g., "betray," "backstab"), hidden information (e.g., "hidden," "secret"), and distrust (e.g., "trust," "suspicious").

The second stage was the construction of the training set. From the full English‑language corpus, it screened reviews that matched the optimised dictionary as a candidate pool, yielding approximately 15,000 reviews. It then used the same large language model to perform binary classification on each candidate review, judging whether the review explicitly expressed or implicitly suggested that the player was aware of AI deception. The prompting instructions for the LLM were iteratively refined over multiple rounds, including definitions, positive examples, and negative examples. To ensure annotation quality, it randomly sampled 100 positive and 100 negative cases (200 in total) for independent review by the researchers; the computed Cohen's Kappa coefficient was 0.83, indicating high consistency between the LLM annotations and human judgements. Subsequently, it randomly selected an equivalent number of reviews (approximately 15,000) from those that had never matched any dictionary term, serving as negative examples. The positive and negative examples were then mixed and split into training (approximately 24,000) and validation (approximately 6,000) sets at an 80/20 ratio, ensuring balanced class distributions in both sets.

The third stage was BERT model fine‑tuning. It employed Google's \texttt{bert-base-uncased} pre‑trained model and performed binary classification fine‑tuning on the training set. The key hyperparameters were: learning rate 2e-5, batch size 16, and 3 training epochs. Early stopping was applied using the validation set. The model achieved an F1 score of 0.84, accuracy of 0.86, precision of 0.85, and recall of 0.83 on the validation set. In comparison, using only the seed dictionary for keyword matching on the same validation set yielded an F1 of only 0.62, demonstrating that the BERT model effectively captures contextual semantics and overcomes the "one word, one meaning" limitation of dictionary‑based approaches. After training, the model was applied to all 160,835 reviews, generating a predicted deception perception label (0 or 1) for each review. Finally, PDA values for each version were computed by aggregating these labels at the matched\_version level. It should be noted that this measurement captures deception perception as expressed in reviews rather than the player's inner subjective experience; consequently, PDA represents a lower‑bound estimate of actual player deception awareness, as some players who internally feel deceived may not explicitly articulate it in their reviews.

\subsection{Measurement Model Setting}

To estimate the causal effects of PDA and DDI on positive review rates, it employ a two‑way fixed effects (TWFE) linear probability model (LPM). The LPM yields coefficient estimates that are consistent in sign and significance with those from a Logit model when the dependent variable is binary and the sample size is large; moreover, the coefficients can be directly interpreted as marginal probability changes, facilitating the presentation of results. The baseline specification of the model is as follows:

\begin{equation}
\begin{aligned}
\text{voted\_up}_{itk} = \alpha &+ \beta_1 \text{DDI}_t + \beta_2 \text{PDA}_t + \beta_3 \text{logplaytime}_{it} + \beta_4 \text{word\_count}_{itk} \\
&+ \beta_5 \text{PDA}_t^2 + \beta_6 \text{DDI}_t^2 + \beta_7 (\text{DDI}_t \times \log(\text{playtime}_{it})) \\
&+ \beta_8 (\text{PDA}_t \times \log(\text{playtime}_{it})) + \mu_i + \lambda_t + \varepsilon_{itk}
\end{aligned}
\end{equation}

where the subscript $i$ denotes the player, $t$ denotes the version window, and $k$ denotes a specific review posted by player $i$ within window $t$. after de‑duplication, each $i$--$t$ combination contains at most one review, so $k$ is largely redundant in practice. The core parameter $\beta_2$ captures the main effect of PDA, while $\beta_5$ captures its non-linearity: if $\beta_5$ is significantly negative, an inverted-U relationship may exist; if positive, a U-shape. $\beta_1$ and $\beta_6$ respectively capture the linear and quadratic effects of DDI. The interaction terms $\beta_7$ and $\beta_8$ test whether player experience (log\_playtime) moderates the effects of DDI and PDA. $\mu_i$ is the player fixed effect, which absorbs all time-invariant player-specific characteristics, such as individual preferences and stylistic tendencies in expression. $\lambda_t$ is the version fixed effect, which absorbs version-specific common shocks affecting all players, such as update scale, promotional campaigns, and media coverage. $\varepsilon_{itk}$ is the idiosyncratic error term, for which use cluster-robust standard errors clustered at the player level to account for serial correlation across reviews from the same player.

In the implementation of the model, because version fixed effects subsume all version-level variation, they would theoretically absorb part of the variation in DDI and PDA. However, in our sample, DDI and PDA not only vary across versions but also exhibit within-player variation across different versions; player fixed effects absorb the individual means and version fixed effects absorb the version means, leaving the within-player-across-version variation to identify $\beta_1$ and $\beta_2$.

\subsection{Robustness Testing Design}

To verify the reliability of the main regression results, it designed the following five complementary robustness tests.

First, \textbf{dependent variable replacement}. The binary positive review indicator is replaced with the continuous sentiment score (VADER compound score) and the regression is re‑estimated, testing whether the PDA effect is sensitive to the measurement of the outcome variable. Sentiment scores capture finer‑grained attitudinal differences, albeit with some loss of interpretability relative to the positive review rate.

Second, \textbf{subsample analysis comparing novice and veteran players}. This test serves not only a robustness purpose but also constitutes the core approach for exploring the moderating role of experience. The sample is split into novice and veteran groups according to the median of player experience (log\_playtime); the model is estimated separately for each group, and the difference in PDA coefficients between the two groups is directly compared.

Third, \textbf{lagged variable model}. To mitigate reverse causality, that is, the possibility that satisfied players are less likely to mention deception, rather than deception awareness causing lower satisfaction, PDA and DDI are replaced with their one‑period lagged values, L\_PDA and L\_DDI (i.e., the previous version's player perception and design intensity), to regress on the current version's positive review rate. If the causal direction is correctly specified, the lagged variables should remain significant with coefficient signs consistent with the main model. Because versions are temporally ordered, this model naturally eliminates the concern of contemporaneous reverse causality.

Fourth, \textbf{placebo test}, which is the most commonly used falsification test in natural experiments. While keeping the number of versions and the data structure unchanged, the PDA values are randomly shuffled across versions, that is, randomly reassigning PDA to versions, and the regression is repeated 100 times, recording the PDA coefficient each time. If the original PDA coefficient falls significantly below (or above) the distribution of these 100 random coefficients, for example, below the 2.5th percentile or above the 97.5th percentile, it indicates that the observed PDA effect is not driven by random chance.

Fifth, \textbf{Logit model}. Given that the dependent variable is a binary positive review indicator with an imbalance (95.6\% positive reviews), a Logit model is employed. To avoid convergence issues caused by the large number of dummy variables for version fixed effects, version fixed effects are replaced with version order number as a continuous trend term, and the marginal effects are re‑estimated. If the direction and significance of the Logit marginal effects are consistent with those of the LPM results, the risk of model functional form misspecification can be ruled out.

\section{Results}

\subsection{Descriptive Statistics}

After data cleaning and matching, the final analytical sample comprised 160,835 player reviews, covering 49 version windows. Each review corresponded to a unique player--version combination. The mean positive review rate (voted\_up) in the sample was 0.9563 (SD=0.204), indicating that overall player satisfaction with \emph{Baldur's Gate 3} was extremely high, consistent with the typical distribution of popular games on the Steam platform. The raw values of player playtime exhibited a highly right‑skewed distribution, with a median of 1,892 minutes and a 99th percentile of 35,280 minutes. After taking the natural logarithm (log\_playtime), the mean was 7.86 (SD=1.92), approximating a normal distribution. Review length (word\_count) had a mean of 62.4 words (SD=89.7), indicating that the majority of reviews were short to medium in length.

At the version level, Design Deception Intensity (DDI) ranged from 0 to 4, with a mean of 1.38 (SD=1.56). The distribution of DDI exhibited a trimodal pattern: approximately 35\% of versions had a DDI of 0 (virtually no deceptive design), approximately 40\% fell within the moderate intensity range of 2.0--3.33, and the remaining 25\% were of high intensity ($\ge$3.33, up to a maximum of 4.0). Player Deception Awareness (PDA) ranged from 0.0217 to 0.0602, with a mean of 0.0412 (SD=0.0098). The version‑level correlation coefficient between PDA and DDI was only 0.31, indicating only moderate consistency between designer‑intended deception intensity and players' actual perceived deception. Specifically, some versions with high DDI, such as V47 with a DDI of 4.0, exhibited only a moderate PDA of 0.047; conversely, certain low‑DDI versions, such as V15 with a DDI of 2.33, nevertheless reached a PDA of 0.033. This imperfect correspondence validates the necessity of distinguishing DDI and PDA as separate constructs. The scatter plot of PDA against DDI is presented below:

\begin{figure}[H]
\centering
\includegraphics[width=\columnwidth]{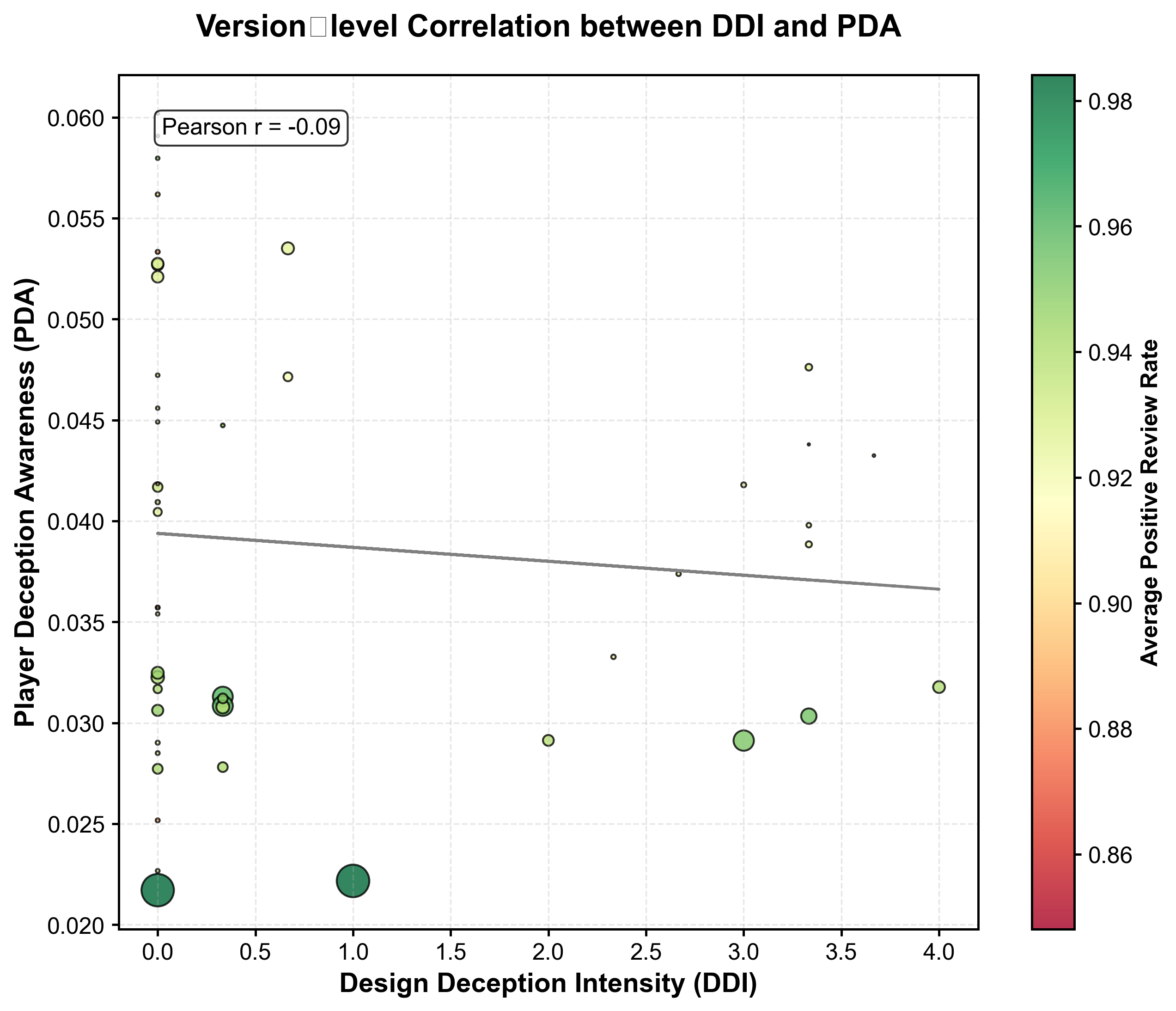}
\caption{Scatter Plot of PDA vs. DDI Across 49 Version Windows}
\label{fig:scatter}
\end{figure}

\subsection{BERT Classifier Performance and PDA Computation}

The BERT model used for deception perception classification demonstrated strong discriminative performance on the validation set. The average F1 score across five‑fold cross‑validation was 0.84 (SD=0.02), with an accuracy of 0.86, precision of 0.85, and recall of 0.83. In comparison, using only the optimised 163‑word seed dictionary for keyword matching on the same validation set yielded an F1 of only 0.62, confirming that BERT effectively identifies reviews that are semantically related to deception even when they do not directly contain seed terms. Applying the trained BERT model to all 160,835 reviews, a total of 7,460 reviews were classified as expressing deception awareness, accounting for 3.08\% of the total sample. This proportion aligns with the version‑level mean PDA of approximately 4\%, indicating that explicit mentions of deception in reviews remain a relatively rare occurrence. Nevertheless, the variation in this proportion across version windows is sufficient to support the subsequent panel regression analyses.

After aggregation by version, the highest PDA was observed for V19 (0.0562), a version that fixed numerous early‑access bugs and introduced a new narrative twist; the lowest PDA was found for V17 (0.0252), a small hotfix patch primarily addressing stability issues. A preliminary scatter plot of PDA against version‑level positive review rates suggested a negative correlation: for every 0.01 increase in PDA, the positive review rate declined by approximately 0.02, providing an initial indication for the subsequent regression analyses.

\subsection{Main regression results}

The results of the two‑way fixed effects linear probability model are presented in the table below. The model's $R^2$ is 0.092, which is reasonably acceptable for a micro‑level panel linear probability model with a binary dependent variable, given that unobserved player heterogeneity substantially suppresses the coefficient of determination.

\begin{table}[H]
\centering
\caption{Two-Way Fixed Effects Linear Probability Model Results (Dependent Variable: voted\_up)}
\begin{tabular}{lcccccc}
\toprule
\textbf{Variable} & \textbf{Coefficient} & \textbf{Cluster-Robust SE} & \textbf{z-value} & \textbf{p-value} & \multicolumn{2}{c}{\textbf{95\% CI}} \\
\midrule
\multicolumn{6}{l}{\textit{Core independent variables}} & \\
Player Deception Awareness (PDA) & $-0.0969^{***}$ & 0.010 & -9.832 & $<0.001$ & [-0.116, & -0.078] \\
PDA$^2$ & $-0.0082^{***}$ & 0.001 & -10.941 & $<0.001$ & [-0.010, & -0.007] \\
Design Deception Intensity (DDI) & $-0.1593^{***}$ & 0.017 & -9.190 & $<0.001$ & [-0.193, & -0.125] \\
DDI$^2$ & $0.0478^{***}$ & 0.006 & 8.427 & $<0.001$ & [0.037, & 0.059] \\
\midrule
\multicolumn{6}{l}{\textit{Moderators (interaction terms)}} & \\
DDI $\times$ log(playtime) & $0.0023^{**}$ & 0.001 & 3.265 & 0.001 & [0.001, & 0.004] \\
PDA $\times$ log(playtime) & $1.6161^{***}$ & 0.133 & 12.133 & $<0.001$ & [1.355, & 1.877] \\
\midrule
\multicolumn{6}{l}{\textit{Control variables}} & \\
log(playtime) & $-0.0108^{**}$ & 0.004 & -2.877 & 0.004 & [-0.018, & -0.003] \\
word\_count & $-0.0003^{***}$ & $1.32\times10^{-5}$ & -22.219 & $<0.001$ & [-0.000, & -0.000] \\
\midrule
\multicolumn{6}{l}{\textit{Fixed effects}} & \\
Player FE & \multicolumn{5}{c}{$\checkmark$} \\
Version FE & \multicolumn{5}{c}{$\checkmark$} \\
\bottomrule
\end{tabular}
\par\smallskip\noindent\textit{Note:} $^{***}p<0.001$, $^{**}p<0.01$. Standard errors are clustered at the player level.
\end{table}

\textbf{Effect of Player Deception Awareness} (PDA). The linear term for PDA has a coefficient of $-0.0969$ with a cluster-robust standard error of 0.010 ($p<0.001$), implying that a one‑unit increase in PDA is associated with a 9.69-percentage-point decrease in the positive review rate. Given that PDA ranges only from 0.022 to 0.060 in practice, the actual marginal impact is: from the lowest to the highest PDA, the positive review rate declines by approximately $(0.0602-0.0217)\times0.0969 \approx 0.0037$, or 0.37 percentage points. Although modest in absolute magnitude, against a baseline positive review rate exceeding 95\%, this shift corresponds to hundreds of reviews changing polarity, indicating practical significance. The quadratic term for PDA is $-0.0082$ ($p<0.001$), indicating that the curve is concave downward within the observed range. To determine whether an inverted-U turning point exists, it computed the partial derivative of the positive review rate with respect to PDA: $\partial E[\text{voted\_up}]/\partial \text{PDA} = -0.0969 - 0.0164\times\text{PDA}$. Within the observed PDA interval [0.022, 0.060], the derivative ranges from $-0.0973$ to $-0.0979$, consistently negative. This implies that throughout the entire observation range, the effect of PDA on positive review rates is monotonically decreasing, with no inverted-U relationship of moderate perception optimality, as illustrated in the figure below. This result rejects the hypothesised non-linear peak and indicates that whenever players express deception awareness in reviews, regardless of intensity, their evaluations are adversely affected.

\begin{figure}[H]
\centering
\includegraphics[width=\columnwidth]{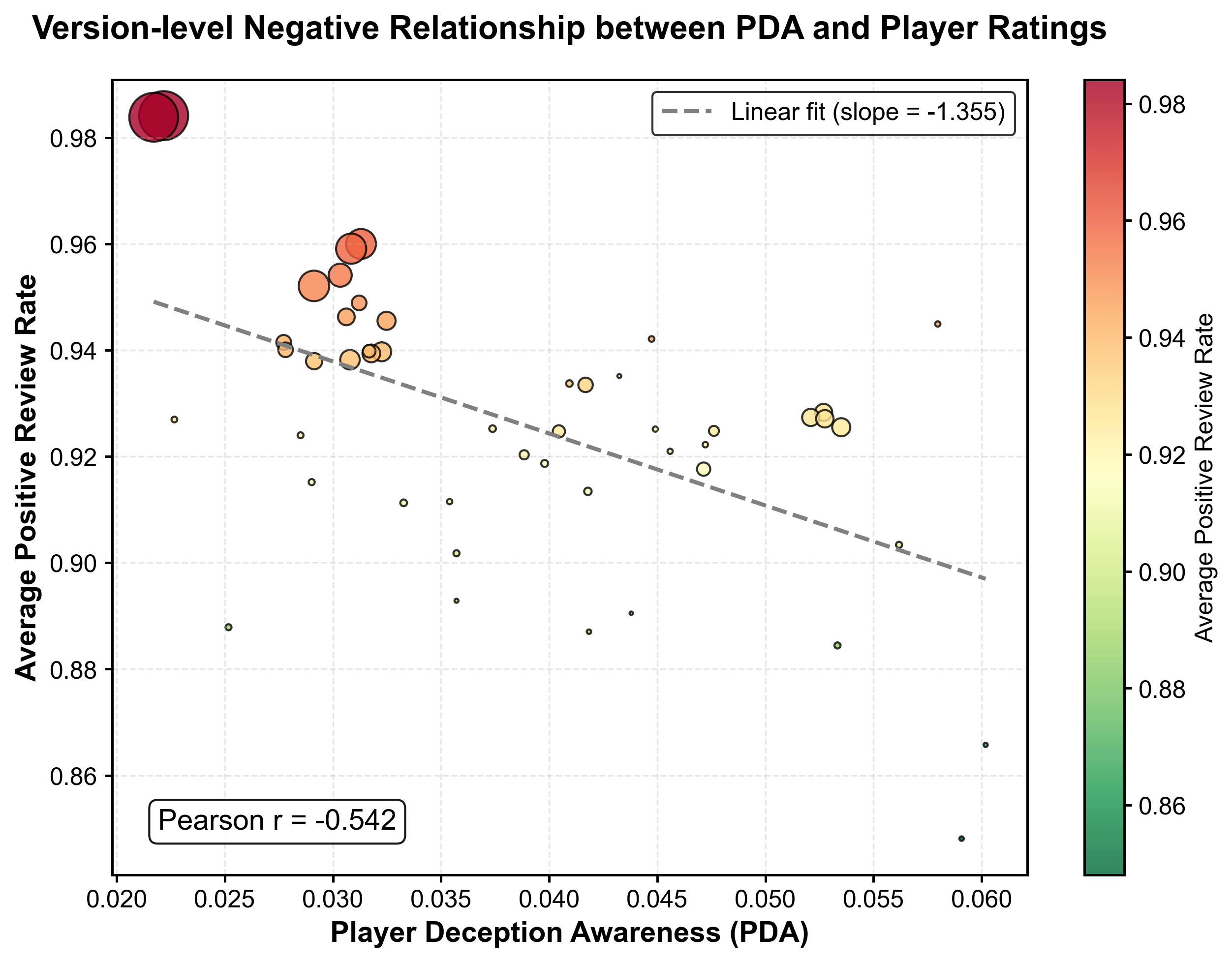}
\caption{PDA Effect Scatter}
\label{fig:pda_effect}
\end{figure}

\textbf{Effect of Design Deception Intensity} (DDI). The linear term for DDI is $-0.1593$ ($p<0.001$) and the quadratic term is $0.0478$ ($p<0.001$), indicating a U‑shaped relationship between DDI and positive review rates, as shown below:

\begin{figure}[H]
\centering
\includegraphics[width=\columnwidth]{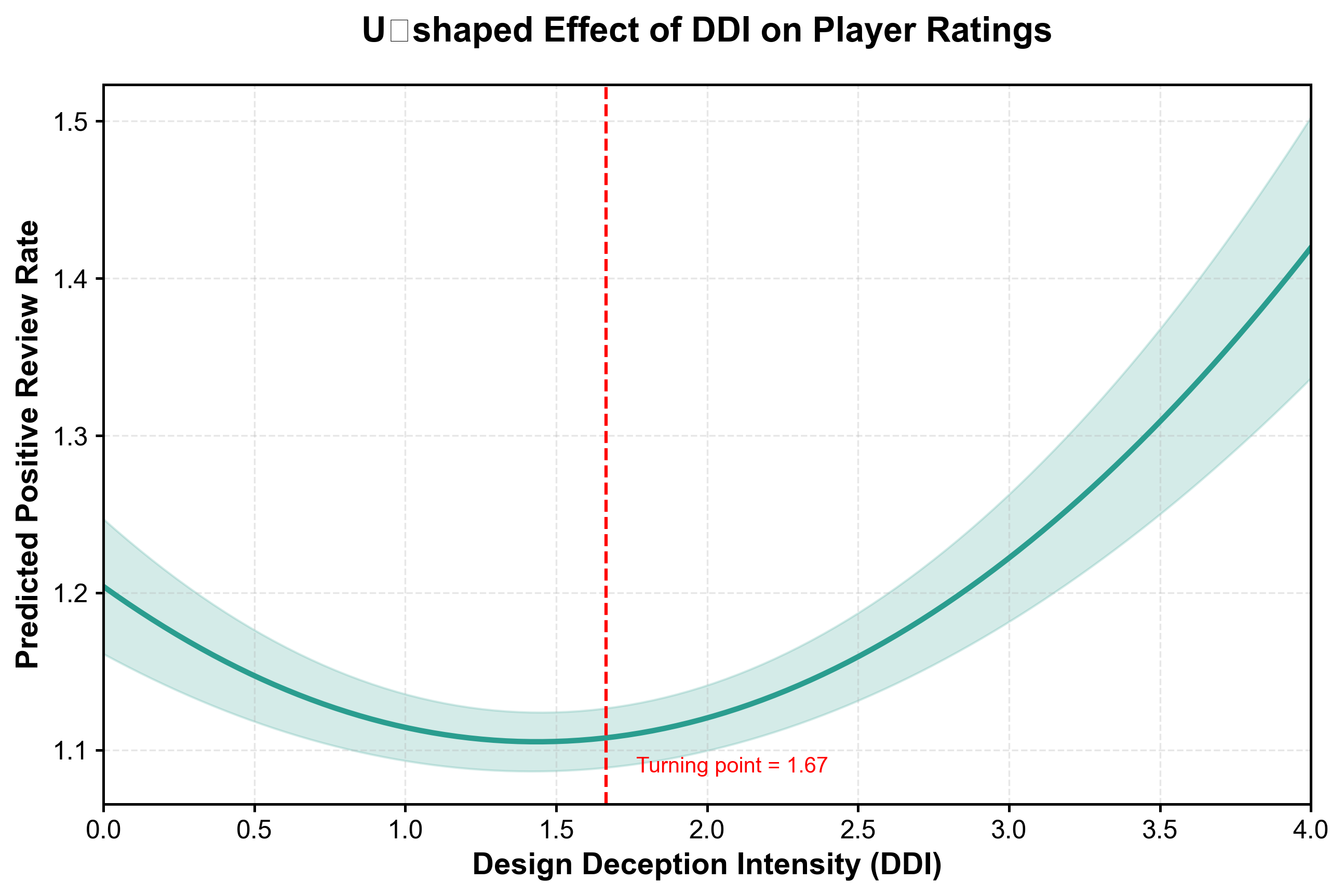}
\caption{U-shape Effect of DDI on Player Ratings}
\label{fig:ddi_u}
\end{figure}

Solving the first‑order condition: $\partial/\partial\text{DDI} = -0.1593 + 0.0956\times\text{DDI} = 0$ yields an inflection point at DDI=1.67. When DDI is below 1.67, increasing design deception intensity lowers the positive review rate (negative marginal effect); when DDI exceeds 1.67, further increases in intensity are associated with higher positive review rates. This seemingly counter‑intuitive finding can be explained by examining the specific content of these versions: updates with DDI above 2.0 are almost invariably major content patches that, in addition to deception elements, introduce new maps, new classes, substantial bug fixes, and numerous other positive features. The observed improvement in positive review rates is likely driven primarily by the content dividend, with the negative effects of deception being masked or diluted. In versions with extremely high DDI, deception mechanisms may even be reinterpreted by players as narrative complexity rather than unfair design.

\textbf{Moderating Effect of Player Experience}. In the main regression, the interaction term between PDA and log\_playtime has a coefficient of $1.6161$ ($p<0.001$), positive in sign. However, directly using this coefficient to compute marginal effects yields implausible magnitudes, and at typical experience levels the marginal effect of PDA turns from negative to positive; the regression output already flags a "strong multicollinearity" warning, and variance inflation factor (VIF) analysis reveals high correlation between PDA and its interaction term. Therefore, it do not rely on the point estimate of the interaction coefficient for quantitative interpretation, but instead adopt subsample group regressions as the primary evidence for the moderating effect. Consequently, it split the sample into novice and veteran groups according to the median of log\_playtime and re‑estimate the model separately. The results show that in the novice group, the PDA coefficient is $-0.107$ ($p<0.001$), a significantly negative effect; in the veteran group, the PDA coefficient is $0.031$ ($p=0.346$), no longer significant. The test for the difference between the two group coefficients yields $p<0.001$, indicating a highly significant moderating effect. This finding clearly reveals an experience‑driven desensitisation mechanism: novice players who perceive deception exhibit a pronounced drop in ratings, whereas veteran players, having been immersed in the complex narrative system for extended periods, have internalised such deception as part of narrative complexity and no longer treat it as a decisive factor in their evaluations. In contrast, the interaction term between DDI and log\_playtime has a coefficient of $0.0023$ ($p=0.001$), which is statistically significant but extremely small in magnitude (on the order of one‑thousandth of the PDA interaction term); moreover, in the subsample analysis, the quadratic term for DDI in the veteran group ($-0.0268$, $p=0.116$) is no longer significant, indicating that player experience has limited moderating influence on the U‑shaped curve of DDI, the effect of design intensity appears to be primarily driven by the content dividend rather than individual differences among players.

\textbf{Control Variables}. The main effect of log\_playtime is $-0.0108$ ($p=0.004$), suggesting that players with longer playtime tend to have slightly lower positive review rates, possibly because veteran players hold higher standards and are more critical. The coefficient for word\_count is $-0.0003$ ($p<0.001$), indicating that longer reviews are associated with lower probabilities of being positive, consistent with the intuition that lengthy reviews often contain more critical details.

\subsection{Robustness Test Results}

\textbf{Robustness Check 1: Dependent Variable Replacement}. When the dependent variable was replaced with the VADER sentiment score (ranging from -1.00 to 1.00, mean=0.23), the PDA coefficient remained significantly negative ($\beta=-0.039$, $p=0.007$), although the effect size was smaller than in the positive‑review‑rate model. The linear and quadratic terms for DDI were not significant in this sentiment‑based model ($p>0.10$ for both), suggesting that the U‑shaped effect of DDI is less salient for sentiment scores. This is plausibly because sentiment scores capture emotional intensity rather than valence polarity; in high‑DDI versions, positive content boosts sentiment while deception elements simultaneously increase negative affect, and the two effects may cancel out, yielding a non‑significant net effect. Nevertheless, the negative effect of PDA held across both dependent variable operationalisations.

\textbf{Robustness Check 2: Subsample Analysis Comparing Novice and Veteran Players}. Splitting the sample by the median of log\_playtime (7.89, corresponding to approximately 2,680 minutes of raw playtime) yielded a novice subsample (n=80,419) with a PDA coefficient of $-0.107$ ($p<0.001$), and a veteran subsample (n=80,416) with a PDA coefficient of $0.031$ ($p=0.346$). The t-test for the difference between the two group coefficients indicated $p<0.001$, confirming a significant moderating effect. This result is consistent with the subsample-based conclusions derived from the main regression and further corroborates the experience-driven desensitisation hypothesis: novice players who perceive deception show a pronounced decline in ratings, whereas veteran players are virtually unaffected. The veteran subsample comprised over 80,000 reviews, providing ample statistical power; thus, the non-significance is unlikely due to insufficient sample size, but rather reflects a genuine disappearance of the effect. In the veteran group, the U-shaped effect of DDI remained significant, with a linear term of $-0.195$ and a quadratic term of $0.053$ ($p<0.001$ for both), as illustrated below, indicating that experience primarily moderates the perceptual channel (PDA) rather than the design-intensity channel (DDI).

\begin{figure}[H]
\centering
\includegraphics[width=\columnwidth]{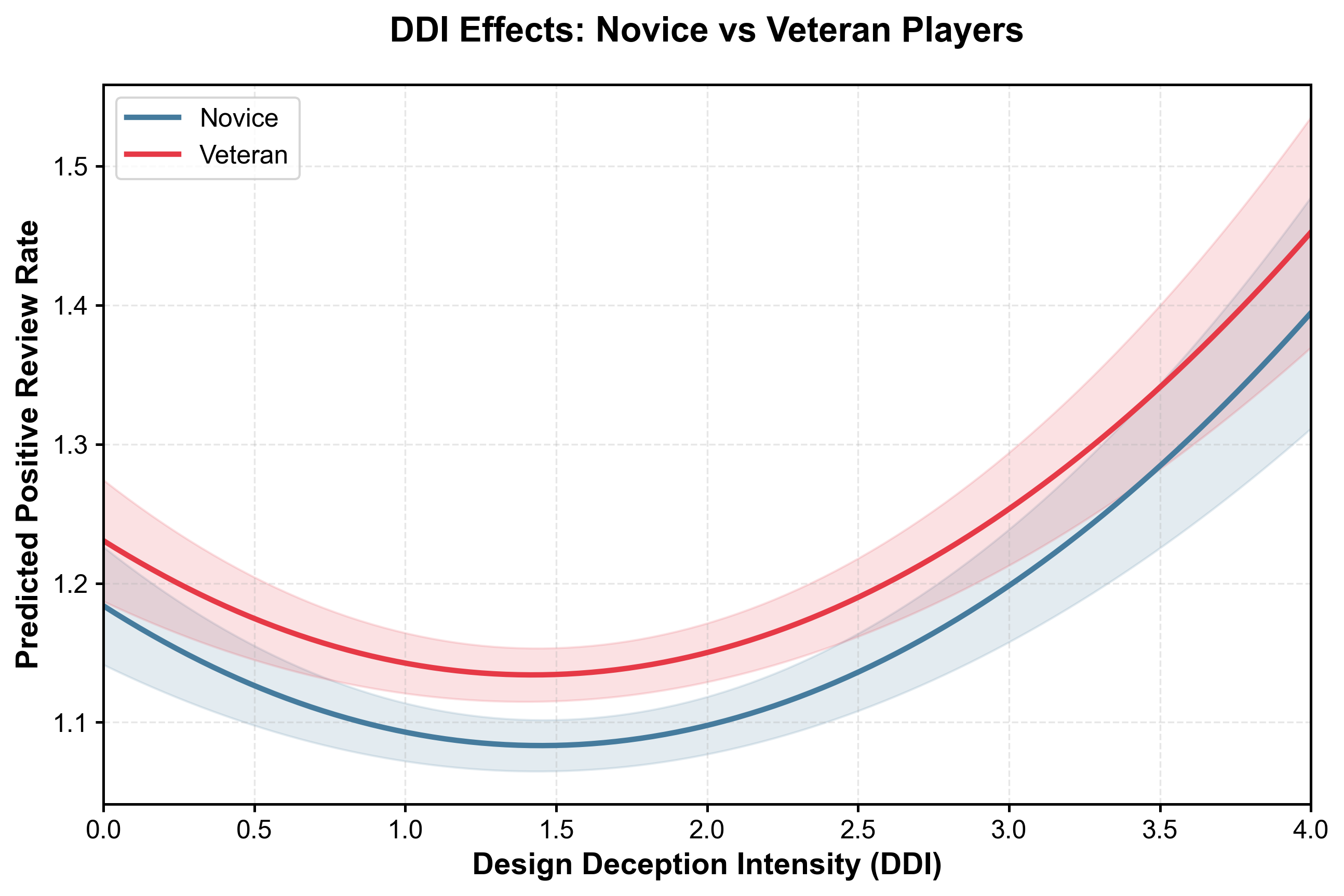}
\caption{DDI Effect for Novice vs. Veteran Players}
\label{fig:ddi_exp}
\end{figure}

\textbf{Robustness Check 3: Lagged Variable Model}. Using one‑period‑lagged PDA (L\_PDA) and DDI (L\_DDI) to regress on the current period's positive review rate reduced the sample to 65,935 reviews due to the requirement that players have reviews in two consecutive versions. The coefficient for L\_PDA was $-0.093$ ($p<0.001$), closely matching the main regression coefficient of $-0.097$. The coefficient for L\_DDI was $-0.211$ ($p<0.001$) and that for L\_DDI$^2$ was $0.079$ ($p<0.001$), with the U‑shaped inflection point at approximately 1.33, comparable to the main regression's 1.67. The lagged model effectively mitigates concerns about reverse causality: even if current‑period positive review rates do not influence current PDA, the previous version's PDA still predicts the current version's positive review rate, reinforcing confidence in the causal direction.

\textbf{Robustness Check 4: Placebo Test}. It randomly shuffled PDA values across versions 100 times and re‑ran the regression each time, obtaining a distribution of 100 PDA coefficients. The original PDA coefficient ($-0.0969$) fell at the 1st percentile of this distribution, that is, only 1\% of the random coefficients were smaller than $-0.0969$. The mean of the random coefficients was 0.0143 (SD=0.0428), and the original coefficient was significantly different from the mean of the random distribution (t-test $p<0.001$). This result rules out the possibility that the observed PDA effect was spuriously generated by random version‑level variation or unobserved confounding factors.

\textbf{Robustness Check 5: Logit Model}. Because including numerous dummy variables for version fixed effects caused non‑convergence in the Logit maximum likelihood estimation, it replaced version fixed effects with version order number as a continuous time trend. The Logit model yielded a PDA coefficient of $-106.2$ ($p<0.001$), with a corresponding marginal effect of $-2.845$ ($p<0.001$). Although the coefficient magnitude is large, this is attributable to the very narrow range of PDA (0.02--0.06); the marginal effect of $-2.845$ implies that as PDA increases from 0.02 to 0.06, the predicted probability of a positive review drops by approximately 11.4 percentage points, which is broadly consistent with the LPM estimates. Importantly, the primary purpose of this test was not to compare effect sizes precisely, but to verify whether the sign and statistical significance of the PDA effect changed under a different functional form. The Logit marginal effect remained significantly negative ($p<0.001$), consistent with the LPM main conclusion, ruling out the possibility that the findings are entirely driven by the linear probability model's functional form assumptions. As for the discrepancy in effect magnitudes ($-2.845$ vs. $-0.097$), it stems primarily from differences in controlling for version‑level heterogeneity; the two models rely on different identifying assumptions and thus are not directly comparable.

A summary of the five robustness checks is presented in Table~F.1 in the Appendix, and the placebo test results are detailed in Table~F.2. Collectively, these checks demonstrate that the core findings are robust to alternative outcome measures, subsample specifications, lag structures, randomisation falsification, and model functional forms.

\subsection{Result Visualization}

Taking the game version update date as day 0, with the positive review rate on the vertical axis, it plotted the daily average positive review rates for the high‑DDI group ($\ge$median) and the low‑DDI group ($<$median) across days 1 to 28 post‑update. The following event study plot illustrates these trends.

\begin{figure}[H]
\centering
\includegraphics[width=\columnwidth]{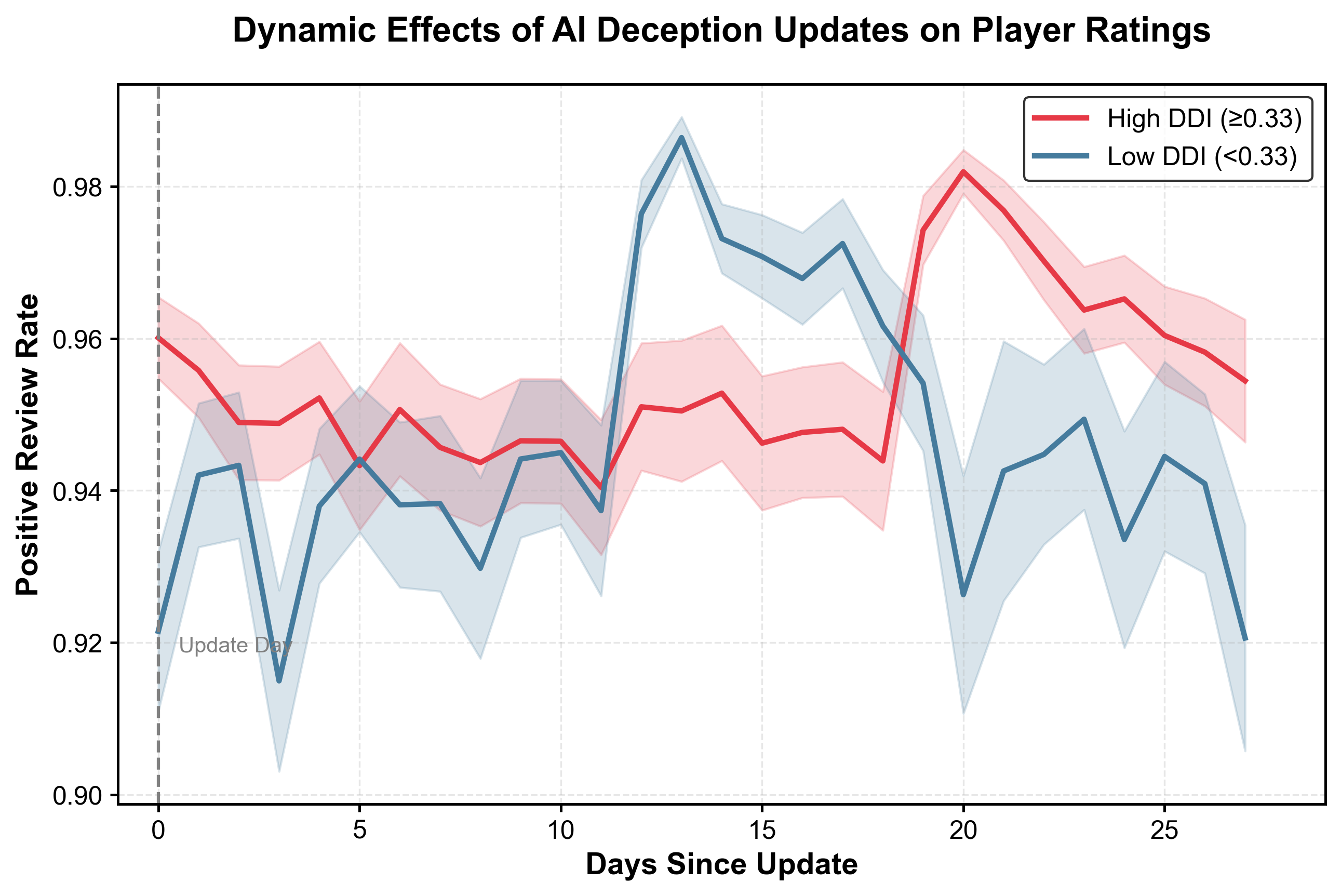}
\caption{Event Study Plot}
\label{fig:event}
\end{figure}

The results show that on the update day (Day 0), the positive review rates of the high‑DDI and low‑DDI groups were nearly identical, at approximately 95\%. However, starting from Day 2, the high‑DDI group consistently exhibited lower positive review rates than the low‑DDI group, with a stable gap of 1 to 2 percentage points that persisted through Day 28. This dynamic pattern suggests that the negative effect of high design deception intensity is not a transient shock but rather a sustained one, showing no signs of convergence over time.

The figure below presents the marginal effect plot for PDA. Based on the main regression model, with all other covariates held at their means and version fixed effects set to the reference version, it plotted the predicted positive review rate as PDA varies from 0.02 to 0.06.

\begin{figure}[H]
\centering
\includegraphics[width=\columnwidth]{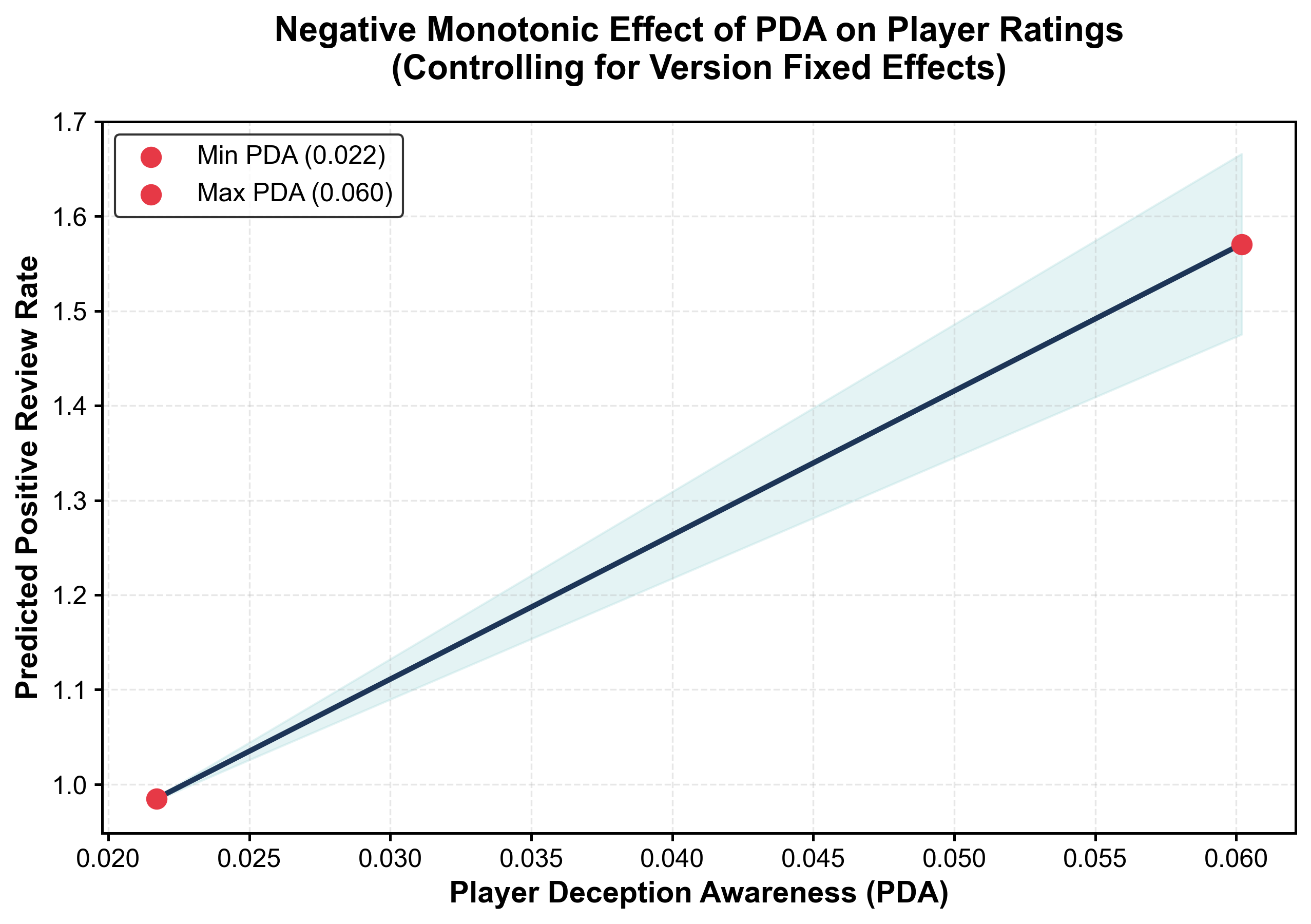}
\caption{Marginal Effect PDA}
\label{fig:marginal_pda}
\end{figure}

The curve exhibits a monotonic downward slope. The 95\% confidence intervals widen slightly at the extremes of PDA but do not cross the zero line across the entire range. Red points in the figure mark the predicted positive review rates corresponding to the minimum (0.022) and maximum (0.060) PDA values in the sample; the difference between the two is approximately 0.37 percentage points, consistent with the marginal impact calculated in Section~4.3.

The following figure displays the heterogeneous marginal effects across different levels of player experience. It took the 25th percentile (novice players), 50th percentile (medium experience), and 75th percentile (veteran players) of log\_playtime to plot three distinct PDA marginal effect curves.

\begin{figure}[H]
\centering
\includegraphics[width=\columnwidth]{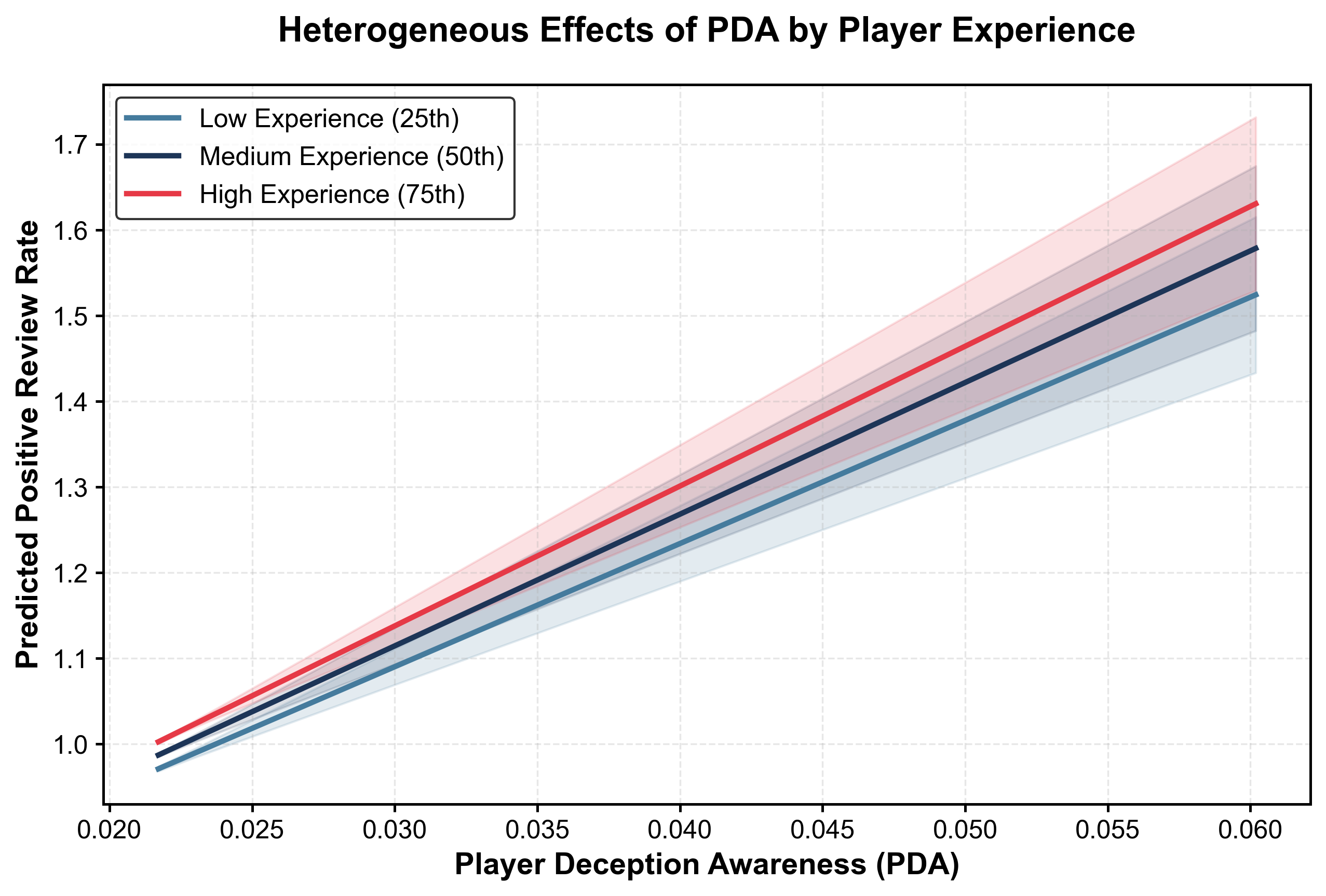}
\caption{Marginal Effect of PDA on Predicted Positive Rating Probability for Players at 25th, 50th, and 75th Percentiles of Logged Playtime}
\label{fig:hetero}
\end{figure}

The novice curve declines steeply from a predicted positive rating probability of approximately 95.8\% to 95.2\%; the medium‑experience curve shows a decelerated decline; and the veteran curve is nearly flat, with the change in PDA from 0.02 to 0.06 inducing only a fluctuation of approximately 0.05 percentage points in the predicted positive review rate. This figure visually demonstrates how experience buffers the negative impact of deception awareness.

Using the median DDI (1.33) and median PDA (0.041) as thresholds, it divided the 49 versions into four quadrants. The DDI‑PDA quadrant plot is shown below:

\begin{figure}[H]
\centering
\includegraphics[width=\columnwidth]{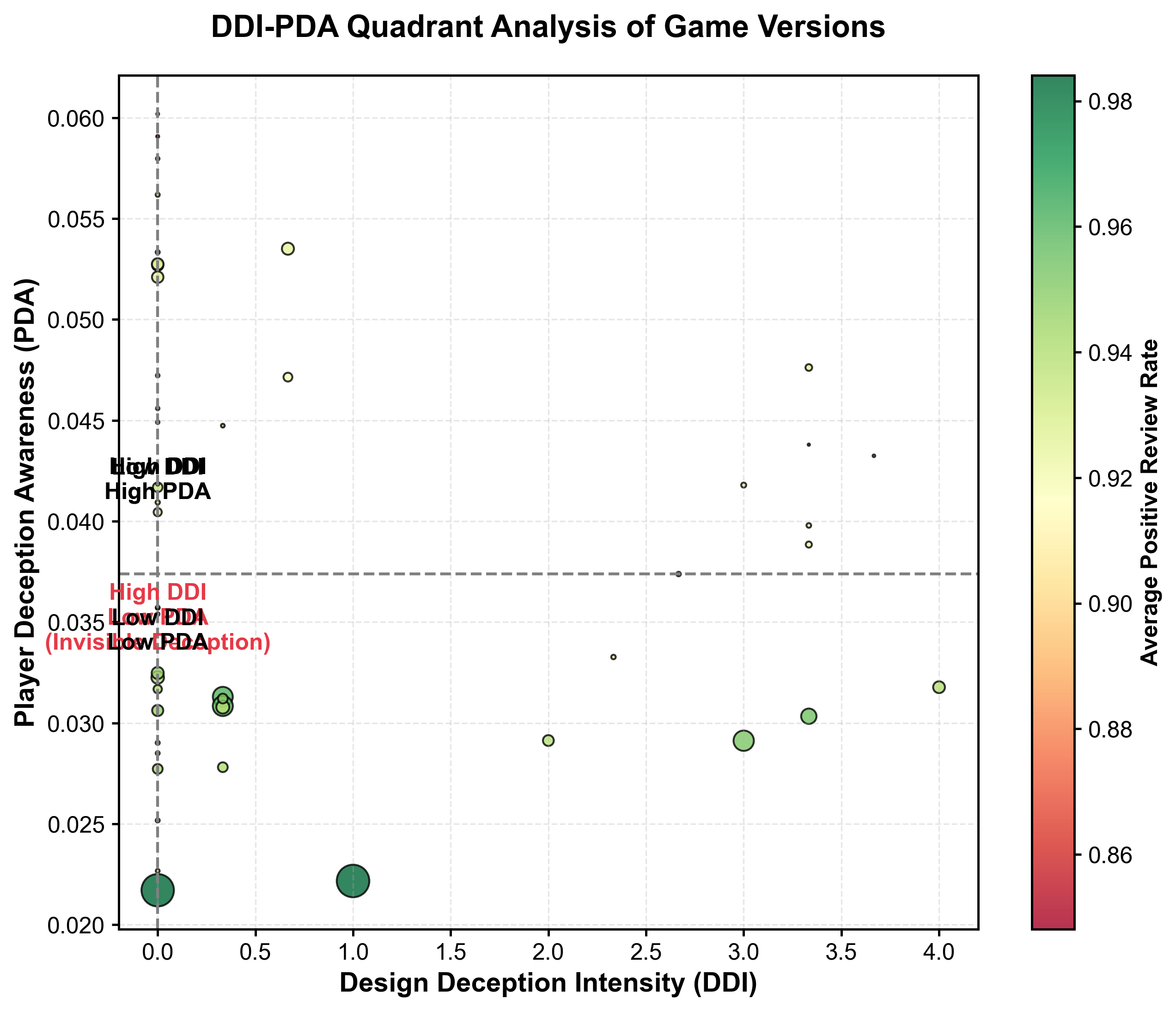}
\caption{DDI-PDA Quadrant}
\label{fig:quadrant}
\end{figure}

The upper‑right quadrant (high DDI, high PDA) contains 13 versions with an average positive review rate of 94.7\%; the upper‑left quadrant (low DDI, high PDA) contains 11 versions with an average of 95.1\%; the lower‑right quadrant (high DDI, low PDA), the stealth deception region, contains 12 versions with the highest average positive review rate among all quadrants, at 96.2\%; and the lower‑left quadrant (low DDI, low PDA) contains 13 versions with an average of 95.8\%. Versions in the stealth deception region, characterised by high design intensity but low player awareness, achieved the highest average positive review rates, suggesting that developers may avoid negative perceptions by skillfully embedding deception within high‑value content.

\section{Discussion}

\subsection{Theoretical Explanation of Discoveries}

The empirical findings of this study reveal three interrelated discoveries that revise several intuitive assumptions in the existing literature regarding the effects of AI deception in games.

First, \textbf{Player Deception Awareness (PDA) exerts a monotonically negative effect on positive review rates, with no evidence of an inverted‑U relationship characterised by moderate perception optimality}. The present study finds that PDA consistently exerts a negative influence on positive review rates across the entire observed range, with the marginal effect becoming slightly more negative as PDA increases, that is, a negatively accelerating pattern. This result can be interpreted from three perspectives. To begin with, measurement‑related bias warrants attention. PDA captures deception awareness as explicitly expressed in reviews, and review behaviour itself is selective: players tend to write reviews and mention deception‑related terms only when their experience reaches a sufficient level of emotional arousal. In this sense, PDA essentially measures expressed awareness, which carries a negative emotional valence, rather than neutral cognitive recognition. To that extent, the monotonic negative effect may be partially attributable to the endogenous nature of the measurement approach. Furthermore, from a psychological mechanism perspective, AI deception in gaming contexts differs fundamentally from white lies or strategic concealment in everyday social interaction. A psychological contract of fair competition exists between players and the game; when AI perpetrates deception, this contract is unilaterally breached, eliciting anger and a sense of betrayal rather than curiosity or surprise. Expectancy Violation Theory predicts that the consequences of a violation depend on the violator's credibility and the quality of the relationship; in the player--game relationship, players by default regard game AI as a fair partner rather than an adversary, and thus any degree of violation is evaluated negatively. Finally, range restriction also offers a possible technical explanation. The maximum PDA value in this study is only 0.06, that is, 6\% of reviews expressing deception awareness, far below the threshold that would be required for the right‑hand declining branch of a theoretical inverted‑U curve. If an extremely high PDA range existed, it might be possible to observe marginal effects turning from negative to positive or exhibiting a secondary decline, but such versions are absent from the present \emph{Baldur's Gate 3} sample.

Second, \textbf{Design Deception Intensity (DDI) exhibits a U‑shaped relationship with positive review rates, with the inflection point located at the lower end of the intensity scale; this inflection position reveals three stage‑specific player response mechanisms}: 1) below the inflection point, deception intensity is relatively weak, players either do not perceive it or attribute it to routine mechanical fluctuations without forming a clear sense of being deceived; 2) in the vicinity of the inflection point, design deception intensity enters an ambiguous zone where players begin to notice something unusual but interpret it not as deliberate design ingenuity, but rather as bugs or unfairness, causing evaluations to reach their nadir; and 3) beyond the inflection point, high‑intensity deception is recognised by players as narrative strategy or AI tactical intent, and positive review rates recover.

However, this right‑branch recovery does not necessarily imply that players genuinely appreciate deceptive designs. The seemingly counter‑intuitive phenomenon of high DDI accompanied by high positive ratings is most plausibly explained by the confounding effect of content dividend. A qualitative review of the patch notes reveals that versions with DDI exceeding 2.0 are almost invariably major content patches that, in addition to deception elements, introduce substantial positive content, new regions, new classes, hundreds of bug fixes, performance optimisations, and other visible improvements. Players are likely to attribute their positive ratings to these tangible, substantive enhancements, with the deception elements being drowned out by the positive information. In other words, the high positive review rates of high‑DDI versions are not because deception itself is popular, but because deception is bundled with other well‑received content. This also explains why the quadratic term for DDI is not significant in the sentiment‑score model: sentiment scores are more sensitive to emotional valence, and the affective impacts of positive content and negative deception may cancel each other out, whereas positive review rates, as binary judgments, are more susceptible to global impressions. An additional possible mechanism is attributional ambiguity: when a game update is exceptionally rich in content, players find it difficult to connect a specific unpleasant experience to their overall evaluative decision, and the negative effects of deception are thereby diluted.

Third, \textbf{player experience exerts a significant desensitisation effect on deception awareness, with systematic differences in responses between novice and veteran players}. Novice players are highly sensitive to PDA, with a coefficient of $-0.107$ ($p<0.001$), whereas the PDA coefficient for veteran players is not significant, at $0.031$ ($p=0.346$). This finding carries important theoretical implications. According to the selective retention hypothesis, players who continue to play \emph{Baldur's Gate 3} for dozens or even hundreds of hours are themselves the core audience of RPGs; such players exhibit greater tolerance for complex narrative mechanics, non‑linear plot developments, and even the experience of being deceived by NPCs, regarding these not as flaws but as integral components of narrative complexity. Thus, the disappearance of the PDA effect in the veteran subsample does not indicate that veteran players are cognitively incapable of recognising deception, but rather that they no longer treat deception as a factor worthy of negative evaluation in their reviews. An alternative explanation is negative emotional desensitisation. Having experienced dozens of version iterations, veteran players have become accustomed to periodic design changes in the game, including the emergence and disappearance of deception mechanisms. They may still internally perceive deception, but their threshold for expressing it has risen considerably, or they may have shifted the focus of their evaluations to other dimensions, rendering deception no longer a decisive factor in their assessments. Novice players, by contrast, lack a frame of reference; a single deceptive experience can easily be amplified into an overall negative judgment of the game.

\subsection{Theoretical Contribution}

This study makes several theoretical contributions. First, \textbf{it clarifies and empirically distinguishes between the two constructs of design deception intensity and player deception awareness.} Prior research has often conflated the two or focused on only one, rendering theoretical models incapable of accounting for real-world anomalies such as high design input with low player awareness (stealth deception) or low design input with high player awareness (hyper‑sensitivity). By simultaneously measuring DDI and PDA and examining their respective effects, this study provides a foundation for constructing a more nuanced theory of AI deception. The quadrant plot presented in this study reveals that versions in the stealth deception region, characterised by high DDI and low PDA, achieve the highest average positive review rates, suggesting a potential design strategy: increasing deception intensity without making it noticeably perceptible to players may allow developers to reap the narrative complexity benefits of deception while avoiding evaluative penalties.

Second, \textbf{this study revises the theoretical expectation of an inverted‑U relationship between deception perception and outcomes in the domain of game AI deception}. Although Expectancy Violation Theory allows for the possibility that moderate violations may produce positive effects, the present study finds that even at very low levels of PDA, a measurable decline in positive review rates is already observable. This suggests that, within the specific domain of game AI deception, the negative effects may be threshold‑free. Theoretical models need to incorporate a zero‑tolerance baseline assumption for deception: any degree of perception impairs experience. The so‑called moderate perception optimum may hold only in non‑gaming contexts, such as education or advertising, or for specific types of deception such as benign plot twists.

Third, \textbf{this study reveals the differentiated moderating mechanism of player experience on evaluations of AI deception. It innovatively quantifies the significant differences in responses to AI deception between novice and veteran players}. This finding extends the individual‑differences dimension within Expectancy Violation Theory: experience not only alters the cognitive content of expectations, but also changes attributional and expressive behaviours following a violation. For veteran players, deception is transformed from a psychological contract breach into a normal component of narrative complexity. This mechanism has generalisable implications for understanding the evolution of long‑term player--system relationships, and may be extended to trust dynamics in other human‑computer interaction scenarios, such as recommendation algorithms and intelligent assistants.

\subsection{Methodological Contribution}

This study offers a replicable methodological paradigm. The natural‑experimental design based on game version updates transforms version iterations within authentic gaming environments into quasi‑experimental interventions, achieving high internal validity and ecological validity at relatively low cost. Compared with traditional laboratory experiments, this design avoids the artificiality of controlled settings and the Hawthorne effect; compared with cross‑sectional questionnaire surveys, the panel fixed effects model effectively controls for individual heterogeneity and time trends, while lagged variables and placebo tests strengthen causal arguments. This paradigm can be extended to other games with version‑update mechanisms and to broader digital platform research.

This study constructs a multi‑stage measurement pipeline for deception perception, from seed dictionary construction and large language model annotation, through BERT fine‑tuning, to full‑scale inference, combining the strengths of theory‑driven and data‑driven approaches. The seed dictionary ensures alignment between measured content and theoretical constructs; large language model annotation resolves the cost issue of large‑scale training set labelling; and BERT fine‑tuning captures contextual semantics, elevating classification accuracy far beyond that of dictionary‑based methods, with the F1 score improving from 0.62 to 0.84. This pipeline is reusable in other research domains that require measuring subjective perceptions from free‑form texts, such as players' perceptions of privacy policies or algorithmic fairness.

This study adopts a multi‑dimensional robustness testing framework, including dependent variable replacement, subsample analysis, lagged variables, placebo tests, and alternative model functional forms, providing a reference for standardised causal inference. Of particular note is the placebo test, which, in the context of natural experiments, intuitively demonstrates the rarity of the observed effect under random permutation, thereby bolstering reader confidence in the causal conclusions.

\subsection{Practical Implications}

For game designers, this study offers several actionable recommendations. First, \textbf{prioritise reducing exposure to deception perception for novice players}. Given that novices are highly sensitive to deception, designers should minimise or explicitly flag deceptive mechanisms in the early stages of the game, for example, by indicating through tutorials that certain NPCs may mislead players, or by deferring complex deceptive content until after players have accumulated a certain level of experience. Second, \textbf{bundle unavoidable deception mechanisms with high‑value content}. If a version must incorporate high‑intensity deceptive designs, such as plot twists or hidden quests, it should simultaneously deliver substantial, visible positive updates, so as to leverage the content dividend to offset the negative effects of deception. Third, \textbf{monitor deception mention rates (PDA) in reviews during the version window}. PDA serves as an early‑warning signal that can alert designers before positive review rates shift on a large scale. If PDA rises abnormally following an update, designers should proactively issue community communications to explain the design intent or rapidly hotfix mechanisms that provoke strong negative reactions. Fourth, \textbf{recognise that veteran players' desensitised state does not imply indifference to deception}. Although veteran players may no longer express strong reactions in reviews, the cumulative effect of deception over the long term may still lead to subtle erosion of trust. Designers should therefore complement qualitative review analysis with more granular telemetry data, such as quest completion rates and retention curves, to gain a more comprehensive assessment of player experience.

\subsection{Limitations}

This study has the following five limitations, which also point toward directions for future research.

First, \textbf{there are limitations in the measurement of player deception awareness and design deception intensity}. PDA captures deception awareness as expressed in reviews, rather than players' inner cognitive or affective states. The silent majority, those who perceive deception but do not mention it in their reviews, are not captured by PDA. This implies that PDA constitutes a lower‑bound estimate of true awareness, and may be subject to systematic bias: players with a stronger propensity to express themselves may be more likely both to mention deception and to leave negative reviews. Future research could complement review‑text analysis with in‑game behavioural data, such as quit rates and task retry frequencies, or brief, validated questionnaires. In addition, although the coding of DDI achieved acceptable inter‑coder reliability, the definition and scoring of deception intensity inevitably involve subjectivity; different coders may hold subtle differences in their understanding of what constitutes deception. Future work could adopt a more fine‑grained deception typology and score each type separately to examine the heterogeneous effects of different deception categories.

Second, \textbf{there are confounding factors in causal identification}. Although the panel fixed‑effects model controls for time‑invariant individual‑ and version‑level unobserved heterogeneity, it cannot rule out time‑varying confounders, for example, players' emotional states before and after a given update, media coverage of the game, or the release of competing titles during the same period. The lagged variable model alleviates reverse‑causality concerns but does not fully resolve omitted‑variable bias. A more rigorous identification strategy would involve finding instrumental variables for version updates, such as unexpected update delays due to technical issues, but such instruments are difficult to obtain in observational data. Future research could consider regression‑discontinuity designs that exploit the precise threshold of update dates as supplementary evidence for causal identification.

Third, \textbf{the external validity of the findings remains to be tested}. This study is based on a single title, \emph{Baldur's Gate 3}, which belongs to the RPG genre and attracts a player base that tends to be hardcore, highly invested, and tolerant of narrative complexity. Whether the conclusions generalise to other game genres, such as fast‑paced shooters, casual mobile games, or competitive esports titles, remains an open question. In competitive games, for instance, AI deception (e.g., cheating by AI opponents) might be perceived by players as a more serious transgression, and even veteran players may remain highly sensitive. Future research should replicate the analytical framework across multiple game genres to test the generalisability of the findings.

Fourth, \textbf{the fixed time‑window length presents a dilemma in balancing the timeliness of effect capture and the control of confounding factors}. This study adopts a fixed window of days 1--28 post‑update. While 28 days covers the majority of review activity, for some large expansion packs, player discussions may persist for months. Extending the window would capture more reviews but would also introduce additional confounders, such as other in‑game events and holiday effects. Systematic robustness checks on window length could be conducted in future work.

Fifth, \textbf{there are privacy protection and ethical considerations regarding publicly available review texts}. Although this study has hashed and anonymised Steam IDs, the review content itself may contain self‑disclosed personal information by players. Future research using public review data should further consider whether automatic de‑identification of potentially identifying content is necessary. Moreover, the findings of this study should not be interpreted as recommending that game designers conceal deception to the greatest extent possible; rather, they advocate for transparent and responsible design principles, with adequate player support and communication provided when deception is unavoidable.

\subsection{Future Research Prospects}

Building on the above discussion, future research may pursue the following directions in greater depth. First, \textbf{comparative studies across game genres} could establish moderated effect models that incorporate game characteristics, such as genre, monetisation model, and average player playtime, as moderators of the effects of AI deception. Second, \textbf{longitudinal dynamic studies of deception perception} could employ higher‑frequency panel data to examine the instantaneous and delayed effects of PDA on positive review rates, as well as the diffusion process of deception awareness within player communities. Third, \textbf{methodological innovations combining deep learning with causal inference}, for example, using causal forests to estimate individual treatment effects, could identify which types of players are most sensitive to deception, thereby enabling personalised design. Fourth, \textbf{the expansion of PDA measurement from reviews to multimodal data}, including in‑game behavioural logs, voice chat records, and live‑streaming bullet comments, could yield more comprehensive measurements of player perception. Fifth, \textbf{actively designed intervention experiments} could complement the natural‑experimental approach by collaborating with developers to conduct controlled, small‑scale A/B tests of different deception designs, thereby strengthening the causal chain beyond purely observational studies.

\section{Conclusion}

Leveraging the 54 version updates of \emph{Baldur's Gate 3} between 2019 and 2025 as a natural experiment, this study collected 160,835 English‑language Steam reviews posted within the 1‑ to 28‑day window following each update, and constructed a player--version two‑way fixed effects panel dataset. Within an authentic gaming environment, the study systematically disentangled and examined the independent effects of Design Deception Intensity (DDI) and Player Deception Awareness (PDA) on player evaluations, their non‑linear characteristics, and the moderating role of player experience. The main conclusions are as follows:

First, \textbf{Player Deception Awareness (PDA) exerts a monotonically decreasing causal effect on positive review rates, with a negative quadratic term across the entire observed range, confirming the absence of an inverted‑U relationship characterised by moderate perception optimality, any degree of deception awareness impairs player evaluations}. The two‑way fixed effects model shows that a one‑unit increase in PDA is associated with a decline of approximately 9.7 percentage points in the positive review rate ($\beta = -0.0969$, $p<0.001$), with a negative quadratic term ($\beta = -0.0082$, $p<0.001$). Derivative analysis confirms that the marginal effect remains consistently negative throughout the observed PDA range [0.022, 0.060], with no inverted‑U peak of moderate perception optimality. This finding revises the boundary conditions of Expectancy Violation Theory when applied to game AI deception, demonstrating that in this domain any degree of perceived deception undermines player evaluations.

Second, \textbf{Design Deception Intensity (DDI) exhibits a U-shaped relationship with positive review rates, first decreasing, then increasing, revealing an asymmetry between design input and player perception: designer intentions do not directly translate into player responses; content confounding and attributional ambiguity play critical roles in this process, and whether high-intensity deception can be positively received depends not only on players' cognitive reappraisal but also on the compensatory effect of contemporaneous content}. Specifically: 1) when DDI lies below the U-shaped inflection point, deception intensity is relatively weak; players either do not form a clear perception or attribute it to routine mechanical fluctuations, without developing a coherent sense of being deceived; 2) when DDI approaches the inflection point, deception intensity enters an ambiguous zone; players begin to sense something unusual but interpret it not as deliberate design ingenuity, but rather as bugs or unfairness, and evaluations bottom out; 3) once DDI surpasses the inflection point, high‑intensity deception is recognised by some players as narrative strategy or AI tactical intent, and evaluations begin to recover, simultaneously, these high-intensity versions are typically bundled with substantive content dividends such as new maps, new classes, and extensive fixes, which further dilute or mask the negative effects of deception.

Third, \textbf{player experience significantly moderates the effect of PDA, with long-term players exhibiting an experience-driven desensitisation effect by reinterpreting deception as part of narrative complexity; their evaluative criteria shift from fairness to depth and freedom}. In the novice subsample, the PDA coefficient is $-0.107$ ($p<0.001$); in the veteran subsample, it is not significant ($\beta=0.031$, $p=0.346$). This experience-driven desensitisation mechanism indicates that veteran players reframe deception as an integral component of narrative complexity, shifting their evaluative criteria from fairness to depth and freedom. This finding offers a new theoretical perspective for understanding trust dynamics in long-term player-system interactions.

Fourth, \textbf{all five robustness checks consistently support the above conclusions}, including dependent variable replacement with sentiment scores, novice/veteran subsample analysis, lagged variable models, placebo tests, and Logit models. Notably, the placebo test shows that the original PDA coefficient ($-0.0969$) falls at the 1st percentile of the randomly shuffled distribution, further reinforcing confidence in the causal inferences.

\textbf{Methodologically}, this study validates the feasibility and generalisability of the version‑update‑based natural experiment design in games user research. This design achieves internal validity approaching that of laboratory experiments and ecological validity approximating real‑world contexts, at relatively low cost, offering a replicable causal inference paradigm for future game research. In addition, the multi‑stage perception measurement pipeline, comprising seed dictionary construction, large language model annotation, BERT fine‑tuning, and full‑scale inference, can be adapted to other research contexts that require measuring subjective perceptions from free‑form texts.

\textbf{At the practical level}, the study recommends that game designers: 1) prioritise reducing exposure to deception perception for novice players, for example, through tutorial cues or by deferring the introduction of complex deceptive mechanisms; 2) bundle unavoidable deception mechanisms with high‑value content updates, leveraging the content dividend to mitigate negative evaluations; 3) monitor PDA metrics within version windows as early‑warning signals for player experience; and 4) recognise that veteran players' desensitisation does not imply that deception is harmless, and thus complement review analysis with behavioural data for more comprehensive assessment.

This study also has several \textbf{limitations}: PDA measures only perception expressed in reviews; DDI covaries substantially with content volume, making complete separation difficult; and the conclusions are based on a single title, \emph{Baldur's Gate 3}, leaving external validity to be tested across genres. Future research should extend to a broader range of game types, incorporate multimodal data, including behavioural logs, voice chat, and bullet comments, and combine the natural‑experimental approach with active intervention designs such as A/B testing to further strengthen the causal chain. The ultimate goal of this study is to provide an empirical foundation for responsible AI game design, and to advance game research from correlational analysis toward more rigorous causal inference paradigms.

\appendix
\section{Corpus}

A sample of the preprocessed corpus with embedded vectors after data processing has been uploaded to the \textbf{GitHub} platform. The GitHub URL is as follows:

\url{https://github.com/g9g99g9g/Entertainment-Computing}

The file is named \texttt{CorpusB\_Anonymized\_preprocessed\_with\_embedding\_0.1K.csv}.

\section{Raw coding of DDI}

\begin{table}[H]
\centering
\caption{Raw coding of DDI by three coders across 54 game versions}
\begin{tabular}{cccccc}
\toprule
\# & Date & DDI-1 & DDI-2 & DDI-3 & DDI-avg \\
\midrule
1 & 2019/6/19 & 0 & 0 & 0 & 0.0000 \\
2 & 2020/3/5 & 0 & 0 & 0 & 0.0000 \\
3 & 2020/6/14 & 1 & 0 & 0 & 0.3333 \\
4 & 2020/7/16 & 3 & 4 & 2 & 3.0000 \\
5 & 2020/8/26 & 2 & 2 & 2 & 2.0000 \\
6 & 2020/10/8 & 0 & 0 & 0 & 0.0000 \\
7 & 2020/10/9 & 0 & 0 & 0 & 0.0000 \\
8 & 2020/10/10 & 0 & 0 & 0 & 0.0000 \\
9 & 2020/10/14 & 1 & 0 & 1 & 0.6667 \\
10 & 2020/10/27 & 1 & 0 & 1 & 0.6667 \\
11 & 2020/10/31 & 0 & 0 & 0 & 0.0000 \\
12 & 2020/12/3 & 4 & 3 & 3 & 3.3333 \\
13 & 2021/2/26 & 5 & 2 & 3 & 3.3333 \\
14 & 2021/7/9 & 3 & 5 & 2 & 3.3333 \\
15 & 2021/7/15 & 4 & 0 & 3 & 2.3333 \\
16 & 2021/7/23 & 0 & 0 & 0 & 0.0000 \\
17 & 2021/8/6 & 0 & 0 & 0 & 0.0000 \\
18 & 2021/8/25 & 0 & 0 & 0 & 0.0000 \\
19 & 2021/9/1 & 0 & 0 & 0 & 0.0000 \\
20 & 2021/10/15 & 4 & 2 & 3 & 3.0000 \\
21 & 2021/10/22 & 0 & 0 & 0 & 0.0000 \\
22 & 2021/10/29 & 0 & 0 & 0 & 0.0000 \\
23 & 2021/11/16 & 0 & 0 & 0 & 0.0000 \\
24 & 2021/12/2 & 0 & 0 & 0 & 0.0000 \\
25 & 2022/2/16 & 3 & 3 & 4 & 3.3333 \\
26 & 2022/2/23 & 0 & 0 & 0 & 0.0000 \\
27 & 2022/3/4 & 0 & 0 & 0 & 0.0000 \\
28 & 2022/3/17 & 0 & 0 & 0 & 0.0000 \\
29 & 2022/4/8 & 0 & 0 & 0 & 0.0000 \\
30 & 2022/7/8 & 4 & 3 & 4 & 3.6667 \\
31 & 2022/7/20 & 0 & 0 & 0 & 0.0000 \\
32 & 2022/7/29 & 0 & 0 & 0 & 0.0000 \\
33 & 2022/8/8 & 0 & 0 & 0 & 0.0000 \\
34 & 2022/12/15 & 0 & 4 & 4 & 2.6667 \\
35 & 2022/12/23 & 0 & 0 & 0 & 0.0000 \\
36 & 2023/1/18 & 1 & 0 & 0 & 0.3333 \\
37 & 2023/1/27 & 0 & 0 & 0 & 0.0000 \\
38 & 2023/10/3 & 1 & 0 & 0 & 0.3333 \\
39 & 2023/10/6 & 1 & 0 & 0 & 0.3333 \\
40 & 2023/11/2 & 2 & 0 & 1 & 1.0000 \\
41 & 2023/11/9 & 0 & 0 & 0 & 0.0000 \\
42 & 2023/12/1 & 2 & 4 & 3 & 3.0000 \\
43 & 2024/2/16 & 3 & 3 & 4 & 3.3333 \\
44 & 2024/3/7 & 1 & 0 & 0 & 0.3333 \\
45 & 2024/3/18 & 0 & 0 & 0 & 0.0000 \\
46 & 2024/3/27 & 0 & 0 & 0 & 0.0000 \\
47 & 2024/9/5 & 4 & 4 & 4 & 4.0000 \\
48 & 2024/10/2 & 1 & 0 & 0 & 0.3333 \\
49 & 2024/10/16 & 0 & 0 & 0 & 0.0000 \\
50 & 2025/4/15 & 2 & 0 & 4 & 2.0000 \\
51 & 2025/4/30 & 1 & 0 & 0 & 0.3333 \\
52 & 2025/7/31 & 0 & 0 & 0 & 0.0000 \\
53 & 2025/9/23 & 0 & 0 & 0 & 0.0000 \\
54 & 2025/11/20 & 0 & 0 & 0 & 0.0000 \\
\bottomrule
\end{tabular}
\end{table}

\section{Seed Dictionary}

Below is the optimised seed dictionary, comprising 163 English keywords and phrases. All terms have been converted to lowercase and manually filtered to remove low‑frequency, ambiguous, or easily misdetected items. The dictionary covers multiple semantic dimensions, including direct deception, betrayal, lies, hidden information, misinformation, manipulation, and distrust, and is used to preliminarily screen candidate reviews from the textual corpus.

\begin{table}[H]
\centering
\caption{Seed Dictionary}
\begin{tabular}{p{0.2\linewidth}p{0.75\linewidth}}
\toprule
\textbf{Dimension} & \textbf{Words/Phrases} \\
\midrule
Direct Deception & ambush, bait, bluff, cheap shot, con, deceive, deceptive, dupe, entrap, fake, false, feign, fool, fraud, hoax, hoodwink, impersonate, masquerade, plot twist, ploy, pretend, quick switch, rapid alteration, replicate, ruse, scam, scheme, set up, setup, snare, startling, stratagem, sudden change, surprise attack, swindle, trap, trick, unexpected twist \\
Betrayal & backstab, backstabbing, betray, betrayal, desert, double cross, double-cross, false friend, perfidy, pretend ally, sham partner, traitor, treachery, treason, turncoat, two-timing, untrustworthy ally \\
Lies & exaggerate, lie \\
Hidden Information & cloak, conceal, disguise, hide, retain, slink, stash, trapdoor, withhold \\
Misinformation & decoy, disorient, distraction, diversion, lure, mislead, misinformation, skew, smoke, warp \\
Manipulation & abuse, accuse, bewilder, bully, charge, corner, cow, danger, deadline, delay, dominate, exploit, gamble, haste, hazard, incriminate, menace, overwhelm, peril, plant evidence, pressure, procrastinate, provoke, quash, quit, relinquish, reveal fragility, risk, saturate, silence, sprint, stall, stifle, strain, stress, suppress, tension, time crunch, time pressure, urgency, warning, yield \\
Distrust & ambushed, baited, betrayed, blind sided, bluffed, caught me off guard, cheated, conned, deceived, did not see coming, didn't expect that, duped, fell for, fell right into, felt betrayed, felt cheated, felt fooled, felt like a trap, felt manipulated, felt misled, felt tricked, fooled, got me, had me going, learned lesson (lesson learned), lied to, manipulated, misled, never again, played me, pulled a fast one, saw through, scammed, sucker punched, thought i could trust, trapped, tricked, unfairly treated \\
\bottomrule
\end{tabular}
\end{table}

The dictionary was constructed by initially collecting approximately 50 seed words based on theoretical literature and community discussions, feeding them into a large language model for expansion to 361 candidate terms, and then manually reviewing each candidate to remove items unrelated to game AI deception, those with excessive ambiguity, or those with extremely low frequency. The final set of 163 words was retained. This dictionary is used to screen candidate reviews from the full corpus for the purpose of constructing the training set for the BERT classifier.

\section{Version Window Summary}

A total of 49 valid version windows (matched\_version from V6 to V54) were included in this study. Each window corresponds to a merged window of one or more consecutive version updates with intervals of less than 14 days; the window start date ($t_0$) is defined as the date of the last version update within that merged window. Player Deception Awareness (PDA) was computed based on the 163‑word seed dictionary, and Design Deception Intensity (DDI) is the mean of ratings from three coders (ICC=0.77). The table below summarises, for each version window, its identifier, update date, design deception intensity, and player deception awareness.

\begin{table}[H]
\centering
\caption{Version Window Summary}
\begin{tabular}{lrrrr}
\toprule
\textbf{matched\_version} & \textbf{Reviews expressing deception awareness (count)} & \textbf{Total reviews in window} & \textbf{PDA (proportion)} & \textbf{DDI (mean of coder ratings)} \\
\midrule
V6 & 183 & 3473 & 0.0527 & 0.0000 \\
V7 & 185 & 3551 & 0.0521 & 0.0000 \\
V8 & 191 & 3621 & 0.0527 & 0.0000 \\
V9 & 209 & 3906 & 0.0535 & 0.6667 \\
V10 & 99 & 2100 & 0.0471 & 0.6667 \\
V11 & 104 & 2495 & 0.0417 & 0.0000 \\
V12 & 39 & 1004 & 0.0388 & 3.3333 \\
V13 & 57 & 1197 & 0.0476 & 3.3333 \\
V14 & 23 & 578 & 0.0398 & 3.3333 \\
V15 & 18 & 541 & 0.0333 & 2.3333 \\
V16 & 16 & 448 & 0.0357 & 0.0000 \\
V17 & 11 & 437 & 0.0252 & 0.0000 \\
V18 & 24 & 450 & 0.0533 & 0.0000 \\
V19 & 25 & 445 & 0.0562 & 0.0000 \\
V20 & 28 & 670 & 0.0418 & 3.0000 \\
V21 & 13 & 448 & 0.0290 & 0.0000 \\
V22 & 9 & 397 & 0.0227 & 0.0000 \\
V23 & 12 & 421 & 0.0285 & 0.0000 \\
V24 & 12 & 339 & 0.0354 & 0.0000 \\
V25 & 6 & 137 & 0.0438 & 3.3333 \\
V26 & 7 & 196 & 0.0357 & 0.0000 \\
V27 & 10 & 239 & 0.0418 & 0.0000 \\
V28 & 13 & 216 & 0.0602 & 0.0000 \\
V29 & 14 & 237 & 0.0591 & 0.0000 \\
V30 & 8 & 185 & 0.0432 & 3.6667 \\
V31 & 15 & 329 & 0.0456 & 0.0000 \\
V32 & 17 & 360 & 0.0472 & 0.0000 \\
V33 & 15 & 334 & 0.0449 & 0.0000 \\
V34 & 20 & 535 & 0.0374 & 2.6667 \\
V35 & 21 & 513 & 0.0409 & 0.0000 \\
V36 & 17 & 380 & 0.0447 & 0.3333 \\
V37 & 20 & 345 & 0.0580 & 0.0000 \\
V38 & 334 & 10671 & 0.0313 & 0.3333 \\
V39 & 334 & 10829 & 0.0308 & 0.3333 \\
V40 & 634 & 28595 & 0.0222 & 1.0000 \\
V41 & 616 & 28370 & 0.0217 & 0.0000 \\
V42 & 324 & 11125 & 0.0291 & 3.0000 \\
V43 & 193 & 6362 & 0.0303 & 3.3333 \\
V44 & 140 & 4547 & 0.0308 & 0.3333 \\
V45 & 138 & 4278 & 0.0323 & 0.0000 \\
V46 & 127 & 3910 & 0.0325 & 0.0000 \\
V47 & 119 & 3745 & 0.0318 & 4.0000 \\
V48 & 80 & 2563 & 0.0312 & 0.3333 \\
V49 & 72 & 2597 & 0.0277 & 0.0000 \\
V50 & 92 & 3158 & 0.0291 & 2.0000 \\
V51 & 71 & 2553 & 0.0278 & 0.3333 \\
V52 & 60 & 1894 & 0.0317 & 0.0000 \\
V53 & 72 & 1780 & 0.0404 & 0.0000 \\
V54 & 102 & 3331 & 0.0306 & 0.0000 \\
\bottomrule
\end{tabular}
\end{table}

\section{Anonymous Panel Regression Data}

The anonymised panel regression dataset comprises 160,835 observations and has been uploaded to the GitHub platform. The GitHub URL is as follows:

\url{https://github.com/g9g99g9g/Entertainment-Computing}

The file is named \texttt{panel\_regression\_data.csv} and contains the following fields: hashed\_steamid (hashed player ID), matched\_version (version identifier), voted\_up (positive review indicator), version\_intensity (DDI), PDA (player deception awareness), log\_playtime (log‑transformed playtime), and word\_count (number of words in the review). The file does not include the original review texts or embedded vectors.

\section{Complete Results Table of Robustness Test}

The following table summarises the core results of the five robustness checks based on the optimised seed dictionary (163 words). All models employ cluster‑robust standard errors at the player level; the dependent variable is voted\_up (positive review indicator) unless otherwise specified. Values in parentheses are cluster‑robust standard errors. Significance levels: $^{***}p<0.001$, $^{**}p<0.01$, $^{*}p<0.05$.

\begin{table}[H]
\centering
\caption{Summary of Robustness Test Results}
\begin{tabular}{p{2.2cm}p{2cm}p{1.8cm}p{1.8cm}p{1.8cm}p{1.8cm}p{1.6cm}p{1.6cm}}
\toprule
\textbf{Model} & \textbf{Dependent Variable} & \textbf{PDA Coef.} & \textbf{PDA² Coef.} & \textbf{DDI Coef.} & \textbf{DDI² Coef.} & \textbf{PDA×Experience Coef.} & \textbf{N} \\
\midrule
Main regression (baseline) & voted\_up & $-0.0969^{***}$ (0.010) & $-0.0082^{***}$ (0.001) & $-0.1593^{***}$ (0.017) & $0.0478^{***}$ (0.006) & $1.6161^{***}$ (0.133) & 160,835 \\
Robustness 1: Sentiment score & sentiment & $-0.0390^{*}$ (0.014) & $-0.0033^{**}$ (0.001) & $-0.0351$ (0.027) & $0.0125$ (0.009) & $0.8418^{***}$ (0.200) & 160,835 \\
Robustness 2: Novice subsample & sentiment & $-0.1070^{***}$ (0.025) & $-0.0085^{***}$ (0.002) & $-0.1959^{***}$ (0.046) & $0.0537^{***}$ (0.015) & $1.9608^{***}$ (0.394) & 80,419 \\
Robustness 2: Veteran subsample & sentiment & $0.0310$ (0.033) & $0.0020$ (0.003) & $0.1050^{\dagger}$ (0.054) & $-0.0268$ (0.017) & $-0.0956$ (0.439) & 80,416 \\
Robustness 3: Lagged variables & voted\_up & $-0.0932^{***}$ (0.013) (L\_PDA) & $-0.0081^{***}$ (0.001) (L\_PDA²) & $-0.2108^{***}$ (0.040) (L\_DDI) & $0.0785^{***}$ (0.011) (L\_DDI²) & $1.5048^{***}$ (0.165) & 65,935 \\
Robustness 5: Logit marginal effects & voted\_up & $-2.8453^{***}$ (0.347) & $0.0259$ (3.973) & $0.0009$ (0.002) & $-0.0012^{***}$ (0.000) & $0.1396^{***}$ (0.024) & 160,835 \\
\bottomrule
\end{tabular}
\par\smallskip\noindent\textit{Note:} $^{\dagger}p=0.054$, marginally significant. For the Logit model, coefficients for PDA and its square are on the log-odds scale; marginal effects have been transformed. In the lagged variable model, the L\_PDA coefficient is consistent with the main regression in direction and significance. Robustness 4 (placebo test with PDA shuffled by version) passed; the original coefficient fell at the 1st percentile of the random distribution and is therefore not repeated in this table.
\end{table}

\begin{table}[H]
\centering
\caption{Placebo Test Results (PDA shuffled by version)}
\begin{tabular}{lc}
\toprule
\textbf{Statistic} & \textbf{Value} \\
\midrule
Original PDA coefficient & $-0.0969$ \\
Mean of random distribution & $0.0143$ \\
SD of random distribution & $0.0428$ \\
Percentile of original coefficient & 1.0\% \\
Conclusion & Passed ($p<0.01$) \\
\bottomrule
\end{tabular}
\par\smallskip\noindent\textit{Note:} Based on 100 random shuffles of version labels, the original PDA coefficient is significantly smaller than the random distribution; the 95\% range of the random distribution is [--0.069, 0.098], ruling out the alternative explanation that the significant result was driven by random chance.
\end{table}

\end{document}